\pdfoutput=1

\documentclass[runningheads]{llncs}

\usepackage[utf8]{inputenc}
\usepackage[british]{babel}
\usepackage{xcolor,soul}
\usepackage{latexsym}
\usepackage{alltt}
\usepackage{listings}
\usepackage{etoolbox}
\usepackage{url}
\usepackage{amsmath}
\usepackage{mathpartir}
\usepackage{tikz}
\usetikzlibrary[arrows]
\usepackage{fancyhdr}
\usepackage[shortcuts]{extdash}
\usepackage[us,24hr]{datetime}
\usepackage{enumitem}

\usepackage{hyperref}
\usepackage{amssymb}
\usepackage{mathtools}
\usepackage{varwidth}
\usepackage{graphicx}
\usepackage{float}
\usepackage{multirow}
\usepackage[noend]{algorithmic}
\usepackage{mathrsfs}
\usepackage{dsfont}
\usepackage{stmaryrd}
\usepackage{textcomp}
\usepackage{bbm}
\usepackage{verbdef}
\usepackage{xspace}
\usepackage{verbatim}
\usepackage{lipsum}
\usepackage{wrapfig}

\usepackage{enumitem}
\usepackage{pifont}
\usepackage{cite}

\expandafter\patchcmd\csname \string\lstinline\endcsname{%
  \leavevmode
  \bgroup
}{%
  \leavevmode
  \ifmmode\hbox\fi
  \bgroup
}{}{%
  \typeout{Patching of \string\lstinline\space failed!}%
}

\newdimen\arraycolseptemp

\makeatletter
\pgfarrowsdeclare{X}{X}
{
  \pgfutil@tempdima=0.3pt%
  \advance\pgfutil@tempdima by.25\pgflinewidth%
  \pgfutil@tempdimb=5.5\pgfutil@tempdima\advance\pgfutil@tempdimb by.5\pgflinewidth%
  \pgfarrowsleftextend{+-\pgfutil@tempdimb}
  \pgfarrowsrightextend{0pt}
}
{
  \pgfutil@tempdima=0.3pt%
  \advance\pgfutil@tempdima by.25\pgflinewidth%
  \pgfsetdash{}{+0pt}
  \pgfsetroundcap
  \pgfsetmiterjoin
  \pgfpathmoveto{\pgfqpoint{-5.5\pgfutil@tempdima}{-6\pgfutil@tempdima}}
  \pgfpathlineto{\pgfqpoint{5.5\pgfutil@tempdima}{6\pgfutil@tempdima}}
  \pgfpathmoveto{\pgfqpoint{-5.5\pgfutil@tempdima}{6\pgfutil@tempdima}}
  \pgfpathlineto{\pgfqpoint{5.5\pgfutil@tempdima}{-6\pgfutil@tempdima}}
  \pgfusepathqstroke
}
\makeatother

\usepackage{xcolor}
\definecolor[named]{ACMBlue}{cmyk}{1,0.1,0,0.1}
\definecolor[named]{ACMYellow}{cmyk}{0,0.16,1,0}
\definecolor[named]{ACMOrange}{cmyk}{0,0.42,1,0.01}
\definecolor[named]{ACMRed}{cmyk}{0,0.90,0.86,0}
\definecolor[named]{ACMLightBlue}{cmyk}{0.49,0.01,0,0}
\definecolor[named]{ACMGreen}{cmyk}{0.20,0,1,0.19}
\definecolor[named]{ACMPurple}{cmyk}{0.55,1,0,0.15}
\definecolor[named]{ACMDarkBlue}{cmyk}{1,0.58,0,0.21}
\hypersetup{colorlinks,
  linkcolor=ACMDarkBlue,
  citecolor=ACMPurple,
  urlcolor=ACMDarkBlue,
  filecolor=ACMDarkBlue}

\newif\ifrevision
\revisionfalse 

\ifrevision
\fi

\let\llncssubparagraph\subparagraph
\let\subparagraph\paragraph
\usepackage[big,compact]{titlesec}
\let\subparagraph\llncssubparagraph

\definecolor{shadecolor}{gray}{1.00}
\definecolor{ddarkgray}{gray}{0.75}
\definecolor{darkgray}{gray}{0.30}
\definecolor{light-gray}{gray}{0.87}

\newcommand{\mute}[1]{#1}

\newcommand{\agp}[1]{\ifrevision \mute{\textcolor{olive}{(\'{A}lvaro: {#1})}} \fi}

\newcommand{\etc}{\emph{etc}}
\newcommand{\ie}{\emph{i.e.}\xspace}

\newcommand{\eg}{\emph{e.g.}\xspace}

\newcommand{\etal}{\emph{et~al.}\xspace}

\newcommand{\cf}{\textit{cf.}\xspace}
\newcommand{\wrt}{\emph{wrt.}\xspace}

{\unskip\nobreak\hskip 1em plus 1fil\nobreak$\square$
\parfillskip=0pt%
\endtrivlist}

\spnewtheorem*{lm}{Lemma}{\bfseries}{}
\spnewtheorem*{thm}{Theorem}{\bfseries}{}

\newcommand{\set}[1]{\left\{{#1}\right\}}

\newcommand{\rulename}[1]{\textsc{#1}\xspace}
\newcommand{\dom}[1]{\mathsf{dom}({#1})}

\newcommand{\angled}[1]{\langle{#1}\rangle}
\newcommand{\Protocol}{\mathcal{P}}
\newcommand{\lstate}{\delta}
\newcommand{\Lstate}{\Delta}
\newcommand{\gstate}{\sigma}
\newcommand{\Gstate}{\Sigma}
\newcommand{\Nodes}{\textsf{Nodes}}
\newcommand{\step}{\mathcal{S}}
\newcommand{\stepi}{\mathcal{S}_{\text{int}}}
\newcommand{\stepr}{\mathcal{S}_{\text{rcv}}}
\newcommand{\steps}{\mathcal{S}_{\text{snd}}}
\newcommand{\mcont}{\mathit{content}}
\newcommand{\To}{\mathit{to}}
\newcommand{\Active}{\mathit{active}}
\newcommand{\Slot}{\mathit{slot}}
\newcommand{\From}{\mathit{from}}

\newcommand{\mvoc}{\mathcal{M}}

\newcommand{\eqdef}{\triangleq}
\newcommand{\pfun}{\rightharpoonup}
\newcommand{\powerset}[1]{\wp({#1})}

\newcommand{\opstep}[1]{\xRightarrow[~~~]{#1}}
\newcommand{\opstepi}[1]{\xRightarrow[~\text{int}~]{#1}}
\newcommand{\opstepr}[1]{\xRightarrow[~\text{rcv}~]{#1}}
\newcommand{\opsteps}[1]{\xRightarrow[~\text{snd}~]{#1}}
\newcommand{\opstepota}[1]{\xRightarrow[~\text{ota}~]{#1}}

\newcommand{\ms}{\mathsf{ms}}
\newcommand{\beh}[1]{\mathcal{B}_{#1}}
\newcommand{\Nat}{\mathbb{N}}
\newcommand{\aand}{\wedge}

\newcommand{\id}{\mathsf{id}}

\newcommand{\Undefined}{\texttt{undef}}
\newcommand{\true}{\texttt{true}}
\newcommand{\false}{\texttt{false}}

\newcommand{\vP}{\texttt{vP}}
\newcommand{\absvP}{\texttt{abs\verbund vP}}
\newcommand{\ptpb}[1]{\texttt{ptp}\verblbrack #1\verbrbrack}

\newcommand{\absresPb}[1]{\texttt{abs\verbund resP}\verblbrack #1\verbrbrack}

\newcommand{\va}{\texttt{v}}

\newcommand{\vRC}{\texttt{vRC}}

\newcommand{\valsRC}{\texttt{valsRC}}
\newcommand{\absvRC}{\texttt{abs\verbund vRC}}
\newcommand{\absroundRC}{\texttt{abs\verbund roundRC}}
\newcommand{\absvalsRC}{\texttt{abs\verbund valsRC}}
\newcommand{\absresRC}{\texttt{abs\verbund resRC}}

\newcommand{\resR}{\texttt{resR}}

\newcommand{\vRR}{\texttt{vRR}}
\newcommand{\roundRR}{\texttt{roundRR}}
\newcommand{\valsRR}{\texttt{valsRR}}

\newcommand{\absvRR}{\texttt{abs\verbund vRR}}
\newcommand{\absroundRR}{\texttt{abs\verbund roundRR}}
\newcommand{\absvalsRR}{\texttt{abs\verbund valsRR}}

\newcommand{\absresrb}[1]
{\texttt{abs\verbund res\verbund r}\verblbrack #1\verbrbrack}
\newcommand{\absreswb}[1]
{\texttt{abs\verbund res\verbund w}\verblbrack #1\verbrbrack}

\newcommand{\countrb}[1]
{\texttt{count\verbund r}\verblbrack #1\verbrbrack}
\newcommand{\countwb}[1]
{\texttt{count\verbund w}\verblbrack #1\verbrbrack}

\newcommand{\prophrb}[1]
{\texttt{proph\verbund r}\verblbrack #1\verbrbrack}
\newcommand{\prophwb}[1]
{\texttt{proph\verbund w}\verblbrack #1\verbrbrack}

\newcommand{\kay}{\texttt{k}}
\newcommand{\jay}{\texttt{j}}
\newcommand{\ar}{\texttt{r}}
\newcommand{\dblu}{\texttt{w}}
\newcommand{\vzero}{\texttt{v0}}

\newcommand{\res}{\texttt{res}}

\newcommand{\True}{\mathsf{True}}
\newcommand{\False}{\mathsf{False}}

\newcommand{\Asrtn}[1]{{\color{blue} #1}}

\newcommand{\InvP}{\mathit{InvP}}
\newcommand{\AbsP}{\mathit{AbsP}}

\newcommand{\InvRC}{\mathit{InvRC}}
\newcommand{\AbsRC}{\mathit{AbsRC}}

\newcommand{\Inv}{\mathit{Inv}}
\newcommand{\AbsRel}{\mathit{AbsRel}}

\newcommand{\mesg}{\mathit{msg}}

\newcommand{\send}{\mathit{sent}}
\newcommand{\receive}{\mathit{received}}
\newcommand{\prid}{\texttt{pid()}}
\newcommand{\reqRead}{\mathit{reqRE}}
\newcommand{\ackRead}{\mathit{ackRE}}

\newcommand{\reqWrite}{\mathit{reqWR}}
\newcommand{\ackWrite}{\mathit{ackWR}}

\newcommand{\AbstractV}{\mathit{AbsV}}
\newcommand{\AbstractVals}{\mathit{AbsVals}}
\newcommand{\AbstractRound}{\mathit{AbsRound}}
\newcommand{\Read}{\mathit{Read}}
\newcommand{\Val}{\mathit{Val}}
\newcommand{\InvRR}{\mathit{InvRR}}
\newcommand{\AbsRR}{\mathit{AbsRR}}
\newcommand{\CountR}{\mathit{CountR}}
\newcommand{\CountW}{\mathit{CountW}}
\newcommand{\Count}{\mathit{Count}}

\newcommand{\ProphROne}{\mathit{ProphR1}}
\newcommand{\ProphRTwo}{\mathit{ProphR2}}
\newcommand{\ProphWOne}{\mathit{ProphW1}}
\newcommand{\ProphWTwo}{\mathit{ProphW2}}
\newcommand{\ProphWThree}{\mathit{ProphW3}}
\newcommand{\Proph}{\mathit{Proph}}

\newcommand{\mmod}{\mathbin{\mathrm{mod}}}

\newcommand{\vW}{\texttt{vW}}
\newcommand{\kW}{\texttt{kW}}

\newcommand{\maxKW}{\texttt{maxKW}}
\newcommand{\maxV}{\texttt{maxV}}

\newcommand{\canon}[1]{\natural{#1}}

\newcommand{\ii}{\texttt{i}}

\newcommand{\verblbrace}{\mbox{\tt\char`\{}}
\newcommand{\verbrbrace}{\mbox{\tt\char`\}}}
\newcommand{\verblbrack}{\mbox{\tt\char`\[}}
\newcommand{\verbrbrack}{\mbox{\tt\char`\]}}
\newcommand{\verbund}{\mbox{\tt\string_}}

\newcommand{\Acceptor}{\texttt{Acceptor}}
\newcommand{\Reader}{\texttt{Reader}}
\newcommand{\Writer}{\texttt{Writer}}

\newcommand{\Prog}{\mathsf{Prog}}
\newcommand{\acc}{\mathsf{acc}}
\newcommand{\pro}{\mathsf{pro}}
\newcommand{\nod}{\mathsf{nod}}

\newcommand{\mpaxos}{Multi-Paxos\xspace}

\newcommand{\cartp}{\times}
\newcommand{\cartw}{\nabla}
\newcommand{\cartow}{\nabla^*}
\newcommand{\reproc}{\mathsf{receiveAndAct}}
\newcommand{\predota}{\mathcal{P}}

\newcommand{\pid}{\mathtt{pid}}
\newcommand{\role}{\mathtt{role}}

\newcommand{\from}{\mathtt{from}}

\newcommand{\bunch}[1]{\mathit{bunch}({#1})}
\newcommand{\msgs}{\mathit{msgs}}

\title{Paxos Consensus, Deconstructed and Abstracted}

\subtitle{Extended Version}

\author{Álvaro {Garc\'{i}a-P\'{e}rez}\inst{1} \and Alexey Gotsman\inst{1} \and Yuri
  Meshman\inst{1} \and Ilya Sergey\inst{2}}

\institute{IMDEA Software Institute, Spain
  \\
  \email{\{alvaro.garcia.perez,alexey.gotsman,yuri.meshman\}@imdea.org}
  \and University College London, UK
  \\
  \email{i.sergey@ucl.ac.uk}
}

\authorrunning{Garc\'{i}a-P\'{e}rez \etal}

\begin{document}

\maketitle

\begin{abstract}


  Lamport's Paxos algorithm is a classic consensus protocol for state machine
  replication in environments that admit crash failures. Many versions of
  Paxos exploit the protocol's intrinsic properties for the sake of gaining
  better run-time performance, thus widening the gap between the original
  description of the algorithm, which was proven correct, and its real-world
  implementations. In this work, we address the challenge of specifying and
  verifying complex Paxos-based systems by (a) devising composable
  specifications for implementations of Paxos's single-decree version, and (b)
  engineering disciplines to reason about protocol-aware, semantics-preserving
  optimisations to single-decree Paxos. In a nutshell, our approach elaborates
  on the deconstruction of single-decree Paxos by Boichat et al. We provide
  novel non-deterministic specifications for each module in the deconstruction
  and prove that the implementations refine the corresponding specifications,
  such that the proofs of the modules that remain unchanged can be reused
  across different implementations. We further reuse this result and show how
  to obtain a verified implementation of Multi-Paxos from a verified
  implementation of single-decree Paxos, by a series of novel protocol-aware
  transformations of the network semantics, which we prove to be
  behaviour-preserving.

\end{abstract}


\section{Introduction}
\label{sec:introduction}

Consensus algorithms are an essential component of the modern fault-tolerant
deterministic services implemented as message-passing distributed systems.
In such systems, each of the distributed nodes contains a replica of the
system's state (\eg, a database to be accessed by the system's clients), and
certain nodes may propose values for the next state of the system (\eg,
requesting an update in the database). Since any node can crash at any moment,
all the replicas have to keep copies of the state that are consistent with
each other.
To achieve this, at each update to the system, all the non-crashed nodes run
an instance of a \emph{consensus protocol}, uniformly deciding on its outcome.
The safety requirements for consensus can be thus stated as follows: ``only a
single value is decided uniformly by all non-crashed nodes, it never changes
in the future, and the decided value has been proposed by some node
participating in the protocol'' \cite{Lamport01}. \agp{This as a quote from
  \cite{Lamport01}.}


The Paxos algorithm \cite{Lamport:TOPLAS98,Lamport01} is the classic consensus
protocol, and its single-decree version (SD-Paxos for short) allows a set of
distributed nodes to reach an agreement on the outcome of a \emph{single}
update.
Optimisations and modifications to SD-Paxos are common. For instance, the
multi-decree version, often called Multi-Paxos
\cite{Lamport:TOPLAS98,VanRenesse-Altinbuken:ACS15}, considers multiple slots
(\ie, multiple positioned updates) and decides upon a result for \emph{each}
slot, by running a slot-specific instance of an SD-Paxos.
Even though it is customary to think of \mpaxos as of a series of independent
SD-Paxos instances, in reality the implementation features multiple
protocol-aware optimisations, exploiting intrinsic dependencies between
separate single-decree consensus instances to achieve better throughput. To a
great extent, these and other optimisations to the algorithm are pervasive,
and verifying a modified version usually requires to devise a new protocol
definition and a proof from scratch.
New versions are constantly springing (\cf Section~5 of
\cite{VanRenesse-Altinbuken:ACS15} for a comprehensive survey) widening the
gap between the description of the algorithms and their real-world
implementations.

We tackle the challenge of \emph{specifying} and \emph{verifying} these
distributed algorithms by contributing two verification techniques for
consensus protocols.

Our first contribution is a family of composable specifications for Paxos'
core subroutines.
Our starting point is the deconstruction of SD-Paxos by
Boichat~\etal\cite{BDFG01,Boichat-al:SN03}, \agp{I believe Alexey wants to
  cite the technical report from 2001. Both cited to avoid controversy.}
allowing one to consider a distributed consensus instance as a
\emph{shared-memory concurrent program}.
We introduce novel specifications for Boichat~\etal's modules, and let them be
non-deterministic.
This might seem as an unorthodox design choice, as it \emph{weakens} the
specification. To show that our specifications are still \emph{strong enough},
we restore the top-level \emph{deterministic} abstract specification of the
consensus, which is convenient for client-side reasoning.
The weakness introduced by the non-determinism in the specifications
%
%
has been impelled by the need to prove
that the implementations of Paxos' components \emph{refine} the specifications
we have ascribed~\cite{Filipovic-al:TCS10}. \agp{First is the need, then the
  specification, not the other way around. The virtue is in better
  understanding the algorithm.} We prove the refinements modularly via the
Rely/Guarantee reasoning with prophecy variables and explicit linearisation
points~\cite{Vafeiadis:PhD,Herlihy-Wing:TOPLAS90}.
On the other hand, this weakness becomes a virtue when better understanding
the volatile nature of Boichat~\etal's abstractions and of the Paxos
algorithm, which may lead to newer modifications and optimisations.



Our second contribution is a methodology for verifying composite consensus
protocols by reusing the proofs of their constituents, targeting specifically
\mpaxos.
%
%
We do so by distilling protocol-aware system optimisations into a separate
semantic layer and showing how to obtain the realistic \mpaxos implementation
from SD-Paxos by a \emph{series of transformations} to the \emph{network
  semantics} of the system, as long as these transformations preserve the
behaviour observed by clients.
We then provide a family of such transformations along with the formal
conditions allowing one to compose them in a behaviour-preserving way.

We validate our approach for construction of modularly verified consensus
protocols by providing an executable proof-of-concept implementation of
\mpaxos with a high-level shared memory-like interface, obtained via a series
of behaviour-preserving network transformations. The full proofs of lemmas and
theorems from our development, as well as some boilerplate definitions, are
given in the appendices at the end of the paper.

\section{The Single-Decree Paxos Algorithm}
\label{sec:smr-con-pax}

We start with explaining SD-Paxos through an intuitive scenario. In SD-Paxos,
each node in the system can adopt the roles of \emph{proposer} or
\emph{acceptor}, or both. A value is decided when a \emph{quorum} (\ie, a
majority of acceptors) accepts the value proposed by some proposer. Now
consider a system with three nodes N1, N2 and N3, where N1 and N3 are both
proposers and acceptors, and N2 is an acceptor, and assume N1 and N3 propose
values $v_1$ and $v_3$, respectively.

\begin{figure}[t]
\begin{tikzpicture}[yscale=.5]
    \draw (.2,3) node {N1:};
    \draw (.2,1.5) node {N2:};
    \draw (.2,0) node {N3:};

    \path[gray, -] (.5,3) edge (11.9,3);
    \path[gray, -] (.5,1.5) edge (11.9,1.5);
    \path[gray, -] (.5,0) edge (11.9,0);

    \draw[->] (.5,3) to[curve to, out=90, in=90, distance=8, <->] (.7,3)
    node[above right] {{\scriptsize \texttt{P1A(1)}}};
    \draw[->] (.5,3) to (.7,1.5) node[above right] {{\scriptsize \texttt{P1A(1)}}};

    \draw (.5,3) to (.65,.75);
    \draw[dotted] (.65,.75) to (.69,.15)
    node[above right] {{\scriptsize \texttt{P1A(1)}}};;

    \draw[->] (1.6,3) to[curve to, out=90, in=90, distance=8, <->] (1.8,3)
    node[above right] {{\scriptsize \texttt{P1B(ok,$\bot$,0)}}};
    \draw[->] (1.6,1.5) node[above right] {{\scriptsize
        \texttt{P1B(ok,$\bot$,0)}}} to (1.8,3);

    \draw[->] (3.5,3) to[curve to, out=90, in=90, distance=8, <->] (3.7,3)
    node[above right] {{\scriptsize \texttt{P2A($v_1$,1)}}};
    \draw[->] (3.5,3)  to (3.7,1.5)
    node[above right] {{\scriptsize \texttt{P2A($v_1$,1)}}};

    \draw (3.5,3) to (3.65,.75);
    \draw[dotted] (3.65,.75) to (3.69,.15)
    node[above right] {{\scriptsize \texttt{P2A($v_1$,1)}}};;

    \draw[->] (5.1,3) to[curve to, out=90, in=90, distance=8, <->] (5.3,3)
    node[above right] {{\scriptsize \texttt{P2B(ok)}}};
    \draw[->] (5.1,1.5) node[above right] {{\scriptsize \texttt{P2B(ok)}}} to (5.3,3);

    \draw[->] (6.1,0) to[curve to, out=-90, in=-90, distance=8, <->] (6.3,0)
    node[below right] {{\scriptsize \texttt{P1A(3)}}};
    \draw[->] (6.1,0) to (6.3,1.5)
    node[below right] {{\scriptsize \texttt{P1A(3)}}};

    \draw (6.1,0) to (6.25,2.25);
    \draw[dotted] (6.25,2.25) to (6.29,2.85)
    node[below right] {{\scriptsize \texttt{P1A(3)}}};

    \draw[->] (7.3,0) to[curve to, out=-90, in=-90, distance=8, <->] (7.5,0)
    node[below right] {{\scriptsize \texttt{P1B(ok,$\bot$,0)}}};
    \draw[->] (7.3,1.5)
    node[below right] {{\scriptsize \texttt{P1B(ok,$v_1$,1)}}} to (7.5,0);

    \draw[->] (9.2,0) to[curve to, out=-90, in=-90, distance=8, <->] (9.4,0)
    node[below right] {{\scriptsize \texttt{P2A($v_1$,3)}}};
    \draw[->] (9.2,0) to (9.4,1.5)
    node[below right] {{\scriptsize \texttt{P2A($v_1$,3)}}};

    \draw (9.2,0) to (9.35,2.25);
    \draw[dotted] (9.35,2.25) to (9.39,2.85)
    node[below right] {{\scriptsize \texttt{P2A($v_1$,3)}}};

    \draw[->] (10.8,0) to[curve to, out=-90, in=-90, distance=8, <->] (11,0)
    node[below right] {{\scriptsize \texttt{P2B(ok)}}};
    \draw[->] (10.8,1.5) node[below right] {{\scriptsize \texttt{P2B(ok)}}} to (11,0);
  \end{tikzpicture}
\vspace{-10pt}
  \caption{A run of SD-Paxos.}
  \label{fig:paxos-run}
\end{figure}
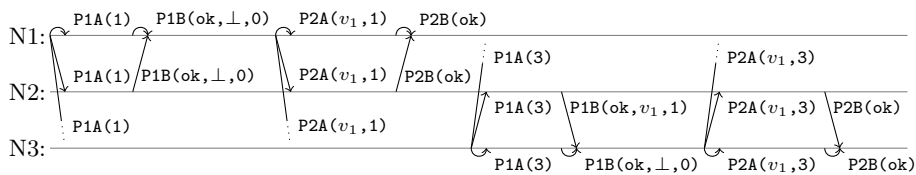

The algorithm works in two phases. In Phase~1, a proposer polls every acceptor
in the system and tries to convince a quorum to promise that they will later
accept its value. If the proposer succeeds in Phase~1 then it moves to
Phase~2, where it requests the acceptors to fulfil their promises in order
to get its value decided. In our example, it would seem in principle possible
that N1 and N3 could respectively convince two different quorums---one
consisting of N1 and N2, and the other consisting of N2 and N3---to go through
both phases and to respectively accept their values. This would happen if the
communication between N1 and N3 gets lost and if N2 successively grants the
promise and accepts the value of N1, and then does the same with N3. This
scenario breaks the safety requirements for consensus because both $v_1$ and
$v_3$---which can be different---would get decided. However, this cannot
happen. Let us explain why.

The way SD-Paxos enforces the safety requirements is by distinguishing each
attempt to decide a value with a unique \emph{round}, where the rounds are
totally ordered. Each acceptor stores its current round, initially the least
one, and only grants a promise to proposers with a round greater or equal than
its current round, at which moment the acceptor switches to the proposer's
round. Figure~\ref{fig:paxos-run} depicts a possible run of the
algorithm. Assume that rounds are natural numbers, that the acceptors' current
rounds are initially $0$, and that the nodes N1 and N3 attempt to decide their
values with rounds $1$ and $3$ respectively. In Phase~1, N1 tries to convince
a quorum to switch their current round to $1$ (messages \texttt{P1A(1)}). The
message to N3 gets lost and the quorum consisting of N1 and N2 switches round
and promises to only accept values at a round greater or equal than $1$. Each
acceptor that switches to the proposer's round sends back to the proposer its
stored value and the round at which this value was accepted, or an undefined
value if the acceptor never accepted any value yet (messages
\texttt{P1B(ok,\,$\bot$,\,0)}, where $\bot$ denotes a default undefined value). 
After Phase~1, N1 picks as a candidate value the one accepted at the greatest
round from those returned by the acceptors in the quorum, or its proposed
value if all acceptors returned an undefined value. In our case, N1 picks its
value $v_1$. In Phase~2, N1 requests the acceptors to accept the candidate
value $v_1$ at round~$1$ (messages \texttt{P2A($v_1$,\,1)}). The message to N3
gets lost, and N1 and N2 accept value $v_1$, which gets decided (messages
\texttt{P2B(ok)}).


Now N3 goes through Phase~1 with round $3$ (messages \texttt{P1A(3)}). Both N2
and N3 switch to round $3$. N2 answers N3 with its stored value $v_1$ and with
the round $1$ at which $v_1$ was accepted (message
\texttt{P1B(ok,\,$v_1$,\,1)}), and N3 answers itself with an undefined value,
as it has never accepted any value yet (message
\texttt{P1B(ok,\,$\bot$,\,0)}). This way, if some value has been already
decided upon, \emph{any} proposer that convinces a quorum to switch to its
round would receive the decided value from some of the acceptors in the quorum
(recall that two quorums have a non-empty intersection). That is, N3 picks the
$v_1$ returned by N2 as the candidate value, and in Phase~2 it manages that
the quorum N2 and N3 accepts $v_1$ at round $3$ (messages
\texttt{P2A($v_1$,\,3)} and \texttt{P2B(ok)}). N3 succeeds in making a new
decision, but the decided value remains the same, and, therefore, the safety
requirements of a consensus protocol are satisfied.



\section{The Faithful Deconstruction of SD-Paxos}
\label{sec:faithfull-deconstruction}


We now recall the faithfull deconstruction of SD-Paxos in
\cite{BDFG01,Boichat-al:SN03}, which we take as the reference architecture for
the implementations that we aim to verify. We later show how each module of
the deconstruction can be verified separately.

\begin{figure}[t]
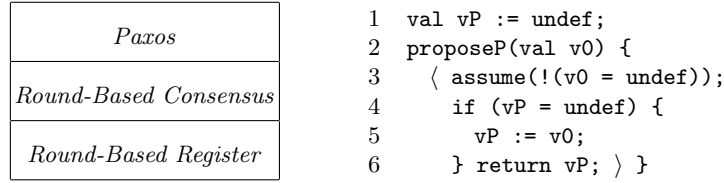

  \begin{minipage}[b]{.55\linewidth}
    \begin{center}
  \renewcommand{\arraystretch}{2}
  \begin{tabular}{| c |}
    \hline
    \emph{Paxos}\\
    \hline
    \emph{Round-Based Consensus}\\
    \hline
    \emph{Round-Based Register}\\
    \hline
  \end{tabular}\renewcommand{\arraystretch}{1}
  \medskip
    \end{center}
  \end{minipage}
  \hfill
  \begin{minipage}[b]{.45\linewidth}
\begin{lstlisting}
val vP := undef;
proposeP(val v0) {
  (@$\langle$@) assume(!(v0 = undef));
    if (vP = undef) {
      vP := v0;
    } return vP; (@$\rangle$@) }
\end{lstlisting}
  \end{minipage}
  \caption{Deconstruction of SD-Paxos (left) and specification of module
    \emph{Paxos} (right).}
  \label{fig:spec-paxos}
\end{figure}

The deconstruction is depicted on the left of Figure~\ref{fig:spec-paxos},
which consists of modules \emph{Paxos}, \emph{Round-Based Consensus} and
\emph{Round-Based Register}. These modules correspond to the ones in Figure~4
of \cite{BDFG01}, with the exception of \emph{Weak Leader Election}. We assume
that a correct process that is trusted by every other correct process always
exists, and omit the details of the leader election. Leaders take the role of
proposers and invoke the interface of \emph{Paxos}. Each module uses the
interface provided by the module below it.

The entry module \emph{Paxos} implements SD-Paxos. Its specification (right of
Figure~\ref{fig:spec-paxos}) keeps a variable \texttt{vP} that stores the
decided value (initially undefined) and provides the operation
\texttt{proposeP} that takes a proposed value \texttt{v0} and returns
\texttt{vP} if some value was already decided, or otherwise it returns
\texttt{v0}. The code of the operation runs \emph{atomically}, which we
emphasise via angle brackets $\angled{\ldots}$. We define this specification
so it meets the safety requirements of a consensus, therefore, any
implementation whose entry point refines this specification will have to meet
the same safety requirements.

In this work we present both specifications and implementations in pseudo-code
for an imperative WHILE-like language with basic arithmetic and primitive
types, where \texttt{val} is some user-defined type for the values decided by
Paxos, and \texttt{undef} is a literal that denotes an undefined value. The
pseudo-code is self-explanatory and we restraint ourselves from giving formal
semantics to it, which could be done in standard fashion if so wished
\cite{Win93}. At any rate, the pseudo-code is ultimately a vehicle for
illustration and we stick to this informal presentation.

The implementation of the modules is depicted in
Figures~\ref{fig:implementation-rb-register-read-write}--\ref{fig:implementation-paxos-rb-consensus}. We
describe the modules following a bottom-up approach, which better fits the
purpose of conveying the connection between the deconstruction and
SD-Paxos. We start with module \emph{Round-Based Register}, which offers
operations \texttt{read} and \texttt{write}
(Figure~\ref{fig:implementation-rb-register-read-write}) and implements the
replicated processes that adopt the role of acceptors
(Figure~\ref{fig:implementation-rb-register-acceptor}). We adapt the
wait-free, crash-stop implementation of \emph{Round-Based Register} in
Figure~5 of \cite{BDFG01} by adding loops for the explicit reception of each
individual message and by counting acknowledgement messages one by
one. Processes are identified by integers from $1$ to $n$, where $n$ is the
number of processes in the system. Proposers and acceptors exchange read and
write requests, and their corresponding acknowledgements and
non\-/acknowledgements. We assume a type \texttt{msg} for messages and let the
message vocabulary to be as follows. Read requests \texttt{[RE,\,k]} carry the
proposer's round \texttt{k}. Write requests \texttt{[WR,\,k,\,v]} carry the
proposer's round \texttt{k} and the proposed value \texttt{v}. Read
acknowledgements \texttt{[ackRE,\,k,\,v,\,k']} carry the proposer's round
\texttt{k}, the acceptor's value \texttt{v}, and the round \texttt{k'} at
which \texttt{v} was accepted. Read non-acknowledgements \texttt{[nackRE,\,k]}
carry the proposer's round \texttt{k}, and so do carry write acknowledgements
\texttt{[ackWR,\,k]} and write non\-/acknowledgements \texttt{[nackWR,\,K]}.

\begin{figure}[t]
\hspace{1.5em}
\begin{minipage}[t]{.55\linewidth}
\begin{lstlisting}
read(int k) {
  int j; val v; int kW; val maxV;
  int maxKW; set of int Q; msg m;
  for (j := 1, j <= (@$n$@), j++)
  { send(j, [RE, k]); }
  maxKW := 0; maxV := undef; Q := {};
  do { (j, m) := receive();
        switch (m) {
          case [ackRE, @k, v, kW]:
            Q := Q (@$\cup$@) {j};
            if (kW >= maxKW)
            { maxKW := kW; maxV := v; }
          case [nackRE, @k]:
            return (false, _);
        } if ((@$|\texttt{Q}|$@) = (@$\lceil$@)(n+1)/2(@$\rceil$@))
          { return (true, maxV); } }
  while (true); }
\end{lstlisting}
\end{minipage}
\hfill
\begin{minipage}[t]{.36\linewidth}
\begin{lstlisting}[firstnumber=18]
write(int k, val vW) {
  int j; set of int Q; msg m;
  for (j := 1, j <= (@$n$@), j++)
  { send(j, [WR, k, vW]); }
  Q := {};
  do { (j, m) := receive();
        switch (m) {
          case [ackWR, @k]:
            Q := Q (@$\cup$@) {j};
          case [nackWR, @k]:
            return false;
        } if ((@$|\texttt{Q}|$@) = (@$\lceil$@)(n+1)/2(@$\rceil$@))
          { return true; } }
  while (true); }
\end{lstlisting}
\end{minipage}
\caption{Implementation of \emph{Round-Based Register} (\texttt{read} and
  \texttt{write}).}
  \label{fig:implementation-rb-register-read-write}
\end{figure}

In the pseudo-code, we use \texttt{\verbund} for a wildcard that could take
any literal value. In the pattern-matching primitives, the literals specify
the pattern against which an expression is being matched, and operator
\texttt{@} turns a variable into a literal with the variable's value. Compare
the \texttt{case\,[ackRE,\,@k,\,v,\,kW]}: in
Figure~\ref{fig:implementation-rb-register-read-write}, where the value of
\texttt{k} specifies the pattern and \texttt{v} and \texttt{kW} get some
values assigned, with the \texttt{case\,[RE,\,k]:} in
Figure~\ref{fig:implementation-rb-register-acceptor}, where \texttt{k} gets
some value assigned.

We assume the network ensures that messages are neither created, modified,
deleted, nor duplicated, and that they are always delivered but with an
arbitrarily large transmission delay.\footnote{We allow creation and
  duplication of \texttt{[RE,\,k]} messages in
  Section~\ref{sec:multi-paxos-via}, where we obtain Multi-Paxos from SD-Paxos
  by a series of transformations of the network semantics.} %
Primitive \texttt{send} takes the destination \texttt{j} and the message
\texttt{m}, and its effect is to send \texttt{m} from the current process to
the process \texttt{j}. Primitive \texttt{receive} takes no arguments, and its
effect is to receive at the current process a message \texttt{m} from origin
\texttt{i}, after which it delivers the pair \texttt{(i,\,m)} of identifier
and message. We assume that \texttt{send} is non-blocking and that
\texttt{receive} blocks and suspends the process until a message is available,
in which case the process awakens and resumes execution.

\begin{figure}[t]
\hspace{1.5em}
\begin{minipage}[t]{1.0\linewidth}
\begin{lstlisting}
process Acceptor(int j) {
  val v := undef; int r := 0; int w := 0;
  start() {
    int i; msg m; int k;
    do { (i, m) := receive();
          switch (m) {
            case [RE, k]:
              if (k < r) { send(i, [nackRE, k]); }
              else { (@$\langle$@) r := k; send(i, [ackRE, k, v, w]); (@$\rangle$@) }
            case [WR, k, vW]:
              if (k < r) { send(i, [nackWR, k]); }
              else { (@$\langle$@) r := k; w := k; v := vW; send(i, [ackWR, k]); (@$\rangle$@) }
          } }
    while (true); } }
\end{lstlisting}
\end{minipage}
\caption{Implementation of \emph{Round-Based Register} (acceptor).}
\label{fig:implementation-rb-register-acceptor}
\end{figure}

Each acceptor (Figure~\ref{fig:implementation-rb-register-acceptor}) keeps a
value \texttt{v}, a current round \texttt{r} (called the \emph{read round}),
and the round \texttt{w} at which the acceptor's value was last accepted
(called the \emph{write round}). Initially, \texttt{v} is \texttt{undef} and
both \texttt{r} and \texttt{w} are $0$.

Phase~1 of SD-Paxos is implemented by operation \texttt{read} on the left of
Figure~\ref{fig:implementation-rb-register-read-write}. When a proposer issues
a \texttt{read}, the operation requests each acceptor's promise to only accept
values at a round greater or equal than \texttt{k} by sending
\texttt{[RE,\,k]} (lines~4--5). When an acceptor receives a \texttt{[RE,\,k]}
(lines~5--7 of Figure~\ref{fig:implementation-rb-register-acceptor}) it
acknowledges the promise depending on its read round. If \texttt{k} is
strictly less than \texttt{r} then the acceptor has already made a promise to
another proposer with greater round and it sends \texttt{[nackRE,\,k]} back
(line~8). Otherwise, the acceptor updates \texttt{r} to \texttt{k} and
acknowledges by sending \texttt{[ackRE,\,k,\,v,\,w]} (line~9). When the
proposer receives an acknowledgement (lines~8--10 of
Figure~\ref{fig:implementation-rb-register-read-write}) it counts
acknowledgements up (line~10) and calculates the greatest write round at which
the acceptors acknowledging so far accepted a value, and stores this value in
\texttt{maxV} (lines~11--12). If a majority of acceptors acknowledged, the
operation succeeds and returns \texttt{(true,\,maxV)}
(lines~15--16). Otherwise, if the proposer received some \texttt{[nackRE,\,k]}
the operation fails, returning \texttt{(false,\,\verbund)} (lines~13--14).

Phase~2 of SD-Paxos is implemented by operation \texttt{write} on the right of
Figure~\ref{fig:implementation-rb-register-read-write}. After having collected
promises from a majority of acceptors, the proposer picks the candidate value
\texttt{vW} and issues a \texttt{write}. The operation requests each acceptor
to accept the candidate value by sending \texttt{[WR,\,k,\,vW]}
(lines~20--21). When an acceptor receives \texttt{[WR,\,k,\,vW]} (line~10 of
Figure~\ref{fig:implementation-rb-register-acceptor}) it accepts the value
depending on its read round. If \texttt{k} is strictly less than \texttt{r},
then the acceptor never promised to accept at such round and it sends
\texttt{[nackWR,\,k]} back (line~11). Otherwise, the acceptor fullfils its
promise and updates both \texttt{w} and \texttt{r} to \texttt{k} and assigns
\texttt{vW} to its value \texttt{v}, and acknowledges by sending
\texttt{[ackWR,\,k]} (line~12). Finally, when the proposer receives an
acknowledgement (lines~23--25 of
Figure~\ref{fig:implementation-rb-register-read-write}) it counts
acknowledgements up (line~26) and checks whether a majority of acceptors
acknowledged, in which case \texttt{vW} is decided and the operation succeeds
and returns \texttt{true} (lines~29--30). Otherwise, if the proposer received
some \texttt{[nackWR,\,k]} the operation fails and returns \texttt{false}
(lines~27--28).\footnote{For the implementation to be correct with our
  shared-memory-concurrency approach, the update of the data in acceptors must
  happen atomically with the sending of acknowledgements in lines~9 and 12 of
  Figure~\ref{fig:implementation-rb-register-acceptor}.}

Next, we describe module \emph{Round-Based Consensus} on the left of
Figure~\ref{fig:implementation-paxos-rb-consensus}. The module offers an
operation \texttt{proposeRC} that takes a round \texttt{k} and a proposed
value \texttt{v0}, and returns a pair \texttt{(res,\,v)} of Boolean and value,
where \texttt{res} informs of the success of the operation and \texttt{v} is
the decided value in case \texttt{res} is \texttt{true}. We have taken the
implementation from Figure~6 in \cite{BDFG01} but adapted to our pseudo-code
conventions. \emph{Round-Based Consensus} carries out Phase~1 and Phase~2 of
SD-Paxos as explained in Section~\ref{sec:smr-con-pax}. The operation
\texttt{proposeRC} calls \texttt{read} (line~3) and if it succeeds then
chooses a candidate value between the proposed value \texttt{v0} or the value
\texttt{v} returned by \texttt{read} (line~5). Then, the operation calls
\texttt{write} with the candidate value and returns \texttt{(true,\,v)} if
\texttt{write} succeeds, or fails and returns \texttt{(false,\,\verbund)}
(line~8) if either the \texttt{read} or the \texttt{write} fails.


\begin{figure}[t]
\hspace{1.5em}
\begin{minipage}[t]{.55\linewidth}
  \begin{lstlisting}
proposeRC(int k, val v0) {
  bool res; val v;
  (res, v) := read(k);
  if (res) {
    if (v = undef) { v := v0; }
    res := write(k, v);
    if (res) { return (true, v); } }
  return (false, _); }
  \end{lstlisting}
\end{minipage}
\hfill
\begin{minipage}[t]{.38\linewidth}
  \begin{lstlisting}
proposeP(val v0) {
  int k; bool res; val v;
  k := pid();
  do { (res, v) :=
          proposeRC(k, v0);
        k := k + (@$n$@);
  } while (!res);
  return v; }
  \end{lstlisting}
\end{minipage}
\caption{Implementation of \emph{Round-Based Consensus} (left) and
  \emph{Paxos} (right)}
  \label{fig:implementation-paxos-rb-consensus}
\end{figure}

Finally, the entry module \emph{Paxos} on the right of
Figure~\ref{fig:implementation-paxos-rb-consensus} offers an operation
\texttt{proposeP} that takes a proposed value \texttt{v0} and returns the
decided value. We assume that the system primitive \texttt{pid()} returns the
process identifier of the current process. We have come up with this
straightforward implementation of operation \texttt{proposeP}, which calls
\texttt{proposeRC} with increasing round until the call succeeds, starting at
a round equal to the process identifier \texttt{pid()} and increasing it by
the number of processes $n$ in each iteration. This guarantees that the round
used in each invocation to \texttt{proposeRC} is~unique.

\subsubsection{The Challenge of Verifying the Deconstruction of Paxos.}

\begin{figure}[t]
  \begin{tikzpicture}[yscale=.5]
    \draw (0,3) node {N1:};
    \draw (0,1.5) node {N2:};
    \draw (0,0) node {N3:};

    \path[thick, |-|] (.5,3) edge node[above=0.1cm] {\texttt{read(1)}}
    node[at end, below=.1cm] {$\bot$} (2.1,3);
    \draw[very thick] (1.3,3.2) edge (1.3,2.8);

    \path[thick, |-|] (2.3,1.5) edge node[above=0.1cm] {\texttt{read(2)}}
    node[at end, below=.1cm] {$v_1$} (9.1,1.5);
    \draw[very thick] (8.4,1.7) edge (8.4,1.3);

    \path[thick, |-|] (2.5,0) edge node[above=0.1cm] {\texttt{read(3)}}
    node[at end, below=.1cm] {$\bot$} (4.1,0);
    \draw[very thick] (3.3,0.2) edge (3.3,-0.2);

    \path[thick, |-|] (4.3,0) edge node[above=0.1cm]
    {\texttt{write(3,$v_3$)}} (6.6,0);
    \draw[very thick] (5.4,0.2) edge (5.4,-0.2);

    \path[thick, |-X] (6.8,3) edge node[above=0.1cm] {\texttt{write(1,$v_1$)}}
    (8.9,3);
    \draw[very thick] (7.8,3.2) edge (7.8,2.8);

    \path[dashed, thick, ->] (7.8,2.8) edge (8.4,1.7);

    \path[thick, |-X] (9.3,1.5) edge node[above=0.1cm]
    {\texttt{write(2,$v_1$)}} (11.5,1.5);
    \draw[very thick] (10.4,1.7) edge (10.4,1.3);
  \end{tikzpicture}

  \smallskip
  \hrule
  \smallskip

  \begin{tikzpicture}[yscale=.5]
    \draw (0,3) node {N1:};
    \draw (0,1.5) node {N2:};
    \draw (0,0) node {N3:};

    \path[thick, |-|] (.5,3) edge node[above=0.1cm] {\texttt{read(1)}}
    node[at end, below=.1cm] {$\bot$} (2.1,3);
    \draw[very thick] (1.3,3.2) edge (1.3,2.8);

    \path[thick, |-X] (2.3,3) edge node[above=0.1cm] {\texttt{write(1,$v_1$)}}
    (9.1,3);
    \draw[very thick] (2.6,3.2) edge (2.6,2.8);

    \path[thick, |-|] (2.6,1.5) edge node[above=0.1cm] {\texttt{read(2)}}
    node[at end, below=.1cm] {$v_1$} (4.7,1.5);
    \draw[very thick] (3,1.7) edge (3,1.3);

    \path[dashed, thick, ->] (2.6,2.8) edge (3,1.7);

    \path[thick, |-|] (4.9,0) edge node[above=0.1cm] {\texttt{read(3)}}
    node[at end, below=.1cm] {$\bot$} (6.5,0);
    \draw[very thick] (5.7,0.2) edge (5.7,-0.2);

    \path[thick, |-|] (6.7,0) edge node[above=0.1cm]
    {\texttt{write(3,$v_3$)}} (8.9,0);
    \draw[very thick] (7.8,0.2) edge (7.8,-0.2);

    \path[thick, |-X] (9.3,1.5) edge node[above=0.1cm]
    {\texttt{write(2,$v_1$)}} (11.5,1.5);
    \draw[very thick] (10.4,1.7) edge (10.4,1.3);
  \end{tikzpicture}

  \caption{Two histories in which a failing \texttt{write} contaminates some
    acceptor.}
  \label{fig:write-contaminates}
\end{figure}
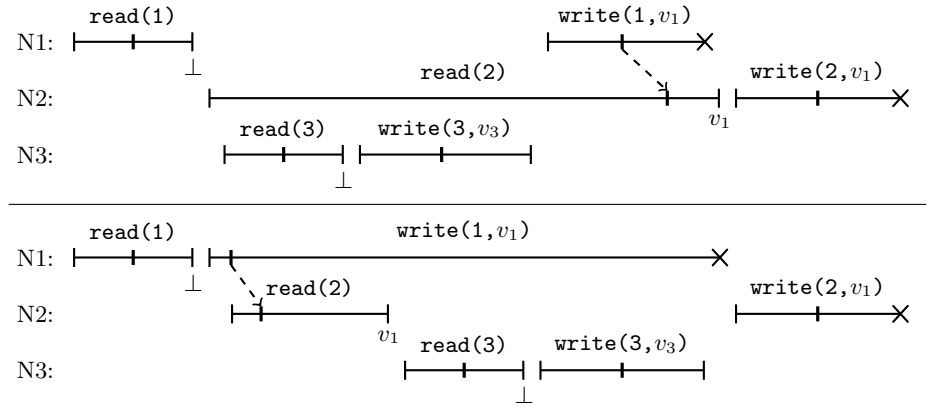

Verifying each module of the deconstruction separately is cumbersome because
of the distributed character of the algorithm and the nature of a
linearisation proof. A process may not be aware of the information that will
flow from itself to other processes, but this future information flow may
dictate whether some operation has to be linearised at the
present. Figure~\ref{fig:write-contaminates} illustrates this challenge.

Let N1, N2 and N3 adopt both the roles of acceptors and proposers, which
propose values $v_1$, $v_2$ and $v_3$ with rounds $1$, $2$ and $3$
respectively. Consider the history on the top of the figure. N2 issues a read
with round $2$ and gets acknowledgements from all but one acceptors in a
quorum. (Let us call this one acceptor A.)  None of these acceptors have
accepted anything yet and they all return $\bot$ as the last accepted value at
round $0$. In parallel, N3 issues a read with round $3$ (third line in the
figure) and gets acknowledgements from a quorum in which A does not
occur. This read succeeds as well and returns \texttt{(true,\,undef)}. Then N3
issues a write with round $3$ and value $v_3$. Again, it gets acknowledgements
from a quorum in which A does not occur, and the write succeeds deciding value
$v_3$ and returns \texttt{true}. Later on, and in real time order with the
write by N3 but in parallel with the read by N2, node N1 issues a write with
round $1$ and value $v_1$ (first line in the figure). This write is to fail
because the value $v_3$ was already decided with round $3$. However, the write
manages to ``contaminate'' acceptor A with value $v_1$, which now acknowledges
N2 and sends $v_1$ as its last accepted value at round $1$. Now N2 has gotten
acknowledgements from a quorum, and since the other acceptors in the quorum
returned $0$ as the round of their last accepted value, the read will catch
value $v_1$ accepted at round $1$, and the operation succeeds and returns
\texttt{(true,\,$v_1$)}. This history linearises by moving N2's read after
N1's write, and by respecting the real time order for the rest of the
operations. (The linearisation ought to respect the information flow order
between N1 and N2 as well, \ie, N1 contaminates A with value $v_1$, which is
read by N2.)

In the figure, a segment ending in an $\times$ indicates that the operation
fails. The value returned by a successful read operation is depicted below the
end of the segment. The linearisation points are depicted with a thick
vertical line, and the dashed arrow indicates that two operations are in the
information flow order.

The variation of this scenario on the bottom of
Figure~\ref{fig:write-contaminates} is also possible, where N1's write and
N2's read happen concurrently, but where N2's read is shifted backwards to
happen before in real time order with N3's read and write. Since N1's write
happens before N2's read in the information flow order, then N1's write has to
inexorably linearise before N3's operations, which are the ones that will
``steal'' N1's valid round.


These examples give us three important hints for designing the specifications
of the modules. First, after a decision is committed it is \emph{not enough}
to store only the decided value, since a posterior write may contaminate some
acceptor with a value different from the decided one. Second, a read operation
\emph{may succeed} with some round even if by that time other operation has
already succeeded with a higher round. And third, a write with a valid round
\emph{may fail} if its round will be ``stolen'' by a concurrent operation. The
non-deterministic specifications that we introduce next allow one to model
execution histories as the ones in Figure~\ref{fig:write-contaminates}.



\section{Modularly Verifying SD-Paxos}
\label{sec:nodet-specs}

In this section, we provide non-deterministic specifications for
\emph{Round-Based Consensus} and \emph{Round-Based Register} and show that
each implementation refines its specification~\cite{Filipovic-al:TCS10}. 
To do so, we instrument the implementations of all the modules with
\emph{linearisation-point} annotations and use Rely/Guarantee
reasoning~\cite{Vafeiadis:PhD}.

This time we follow a top-down order and start with the entry module
\emph{Paxos}.

\begin{figure}[t]
\hspace{1.5em}
\begin{minipage}{1.0\linewidth}
  \begin{lstlisting}
(@\hl{(bool $\times$ val) ptp[1..$n$] := undef;}@)
(@\hl{val abs\verbund vP := undef; single bool abs\verbund resP[1..$n$] := undef;}@)
proposeP(val v0) {
  int k; bool res; val v; (@\hl{assume(!(v0 = undef));}@)
  k := pid(); (@\hl{ptp[pid()] := (true, v0);}@)
  do { (@\hl{$\langle$}@) (res, v) := proposeRC(k, v0);
          (@\hl{if (res) \verblbrace}@)
            (@\hl{for (i := 1, i <= $n$, i++) \verblbrace}@)
              (@\hl{if (ptp[i] = (true, v)) \verblbrace~lin(i); ptp[i] := (false, v); \verbrbrace~\verbrbrace}@)
            (@\hl{if (!(v = v0)) \verblbrace~lin(pid()); ptp[pid()] := (false, v0); \verbrbrace~\verbrbrace~$\rangle$}@)
       k := k + (@$n$@); }
  while (!res); return v; }
  \end{lstlisting}
 \end{minipage}
\caption{Instrumented implementation of \emph{Paxos}.}
  \label{fig:single-decree}
\end{figure}

\subsubsection{Module \emph{Paxos}.}
In order to prove that the implementation on the right of
Figure~\ref{fig:implementation-paxos-rb-consensus} refines its specification
on the right of Figure~\ref{fig:spec-paxos}, we introduce the instrumented
implementation in Figure~\ref{fig:single-decree}, which uses the helping
mechanism for external linearisation points of \cite{Liang-Feng:PLDI13}. We
assume that each proposer invokes \texttt{proposeP} with a unique proposed
value. The auxiliary pending thread pool \texttt{ptp[}$n$\texttt{]} is an
array of pairs of Booleans and values of length $n$, where $n$ is the number
of processes in the system. A cell \texttt{ptp[$i$]} containing a pair
\texttt{(true,\,$v$)} signals that the process $i$ proposed value $v$ and the
invocation \texttt{proposeP($v$)} by process $i$ awaits to be linearised. Once
this invocation is linearised, the cell \texttt{ptp[$i$]} is updated to the
pair \texttt{(false,\,$v$)}. A cell \texttt{ptp[$i$]} containing
\texttt{undef} signals that the process $i$ never proposed any value yet. The
array \texttt{abs\verbund resP[$n$]} of Boolean single-assignment variables
stores the abstract result of each proposer's invocation. A
linearisation-point annotation \texttt{lin($i$)} takes a process identifier
$i$ and performs atomically the abstract operation invoked by proposer $i$ and
assigns its result to \texttt{abs\verbund resP[$i$]}. The abstract state is
modelled by variable \texttt{abs\verbund vP}, which corresponds to variable
\texttt{vP} in the specification on the right of
Figure~\ref{fig:spec-paxos}. One invocation of \texttt{proposeP} may help
linearise other invocations as follows. The linearisation point is together
with the invocation to \texttt{proposeRC} (line~6). If \texttt{proposeRC}
committed with some value \texttt{v}, the instrumented implementation traverses
\texttt{ptp} and linearises all the proposers which were proposing value
\texttt{v} (the proposer may linearise itself in this traversal)
(lines~8--9). Then, the current proposer linearises itself if its proposed
value \texttt{v0} is different from \texttt{v} (line~10), and the operation
returns \texttt{v} (line~12). All the annotations and code in lines~6--10 are
executed inside an atomic block, together with the invocation to
\texttt{proposeRC(k,\,v0)}.

\begin{theorem}
  \label{th:paxos}
  The implementation of Paxos on the right of
  Figure~\ref{fig:implementation-paxos-rb-consensus} linearises with respect
  to its specification on the right of Figure~\ref{fig:spec-paxos}.
\end{theorem}



\begin{figure}[t]
  \hspace{1.5em}
  \begin{minipage}[t]{1.0\linewidth}
\begin{lstlisting}
val vRC := undef; int roundRC := 0; set of val valsRC := {};
proposeRC(int k, val v0) {
  (@$\langle$@) val vD := random(); bool b := random();
    assume(!(v0 = undef)); assume(pid() = ((k - 1) mod (@$n$@)) + 1);
    if (vD (@$\in$@) (valsRC (@$\cup$@) {v0})) {
      valsRC := valsRC (@$\cup$@) {vD};
      if (b && (k >= roundRC)) { roundRC := k;
                                    if (vRC = undef) { vRC := vD; }
                                    return (true, vRC); }
      else { return (false, _); } }
    else { return (false, _); } (@$\rangle$@) }
\end{lstlisting}
\end{minipage}

  \hspace{1.5em}
  \begin{minipage}[t]{1.0\linewidth}
\begin{lstlisting}
(@\hl{val abs\verbund vRC := undef; int abs\verbund roundRC := 0;}@)
(@\hl{set of val abs\verbund valsRC := \verblbrace\verbrbrace;}@)
proposeRC(int k, val v0) {
  (@\hl{single (bool $\times$ val) abs\verbund resRC := undef;}@) bool res; val v;
  (@\hl{assume(!(v0 = undef)); assume(pid() = ((k - 1) mod $n$) + 1);}@)
  (@\hl{$\langle$}@) (res, v) := read(k); (@\hl{if (res = false) \verblbrace~linRC(undef, \verbund); \verbrbrace~$\rangle$}@)
  if (res) { if (v = undef) { v := v0; }
              (@\hl{$\langle$}@) res := write(k, v); (@\hl{if (res) \verblbrace~linRC(v, true); \verbrbrace}@)
                                      (@\hl{else \verblbrace~linRC(v, false);~\verbrbrace~$\rangle$}@)
              if (res) { return (true, v); } }
  return (false, _); }
\end{lstlisting}
  \end{minipage}

  \caption{Specification (top) and instrumented implementation (bottom) of
    \emph{Round-Based Consensus}.}
  \label{fig:spec-inst-imp-round-based-consensus}
\end{figure}


\subsubsection{Module \emph{Round-Based Consensus}.}
The top of Figure~\ref{fig:spec-inst-imp-round-based-consensus} shows the
non-deterministic module's specification. Global variable \texttt{vRC} is the
decided value, initially \texttt{undef}. Global variable \texttt{roundRC} is
the highest round at which some value was decided, initially $0$; a global set
of values \texttt{valsRC} (initially empty) contains values that may have been
proposed by proposers. The specification is non-deterministic in that local
value \texttt{vD} and Boolean \texttt{b} are unspecified, which we model by
assigning random values to them. We assume that the current process identifier
is $((\texttt{k}-1)\mmod n)+1$, which is consistent with how rounds are
assigned to each process and incremented in the code of \texttt{proposeP} on
the right of Figure~\ref{fig:implementation-paxos-rb-consensus}. If the
unspecified value \texttt{vD} is neither in the set \texttt{valsRC} nor equal
to \texttt{v0} then the operation returns \texttt{(false,\,\verbund)}
(line~11). This models that the operation fails without contaminating any
acceptor. Otherwise, the operation may contaminate some acceptor and the value
\texttt{vD} is added to the set \texttt{valsRC} (line~6). Now, if the
unspecified Boolean \texttt{b} is false, then the operation returns
\texttt{(false,\,\verbund)} (lines~7 and 10), which models that the round will
be stolen by a posterior operation. Finally, the operation succeeds if
\texttt{k} is greater or equal than \texttt{roundRC} (line~7), and
\texttt{roundRC} and \texttt{vRC} are updated and the operation returns
\texttt{(true,\,vRC)} (lines~7--9).

In order to prove that the implementation in
Figure~\ref{fig:implementation-paxos-rb-consensus} linearises with respect to
the specification on the top of
Figure~\ref{fig:spec-inst-imp-round-based-consensus}, we use the instrumented
implementation on the bottom of the same figure, where the abstract state is
modelled by variables \texttt{abs\verbund vRC}, \texttt{abs\verbund roundRC}
and \texttt{abs\verbund valsRC} in lines~1--2, the local single-assignment
variable \texttt{abs\verbund resRC} stores the result of the abstract
operation, and the linearisation-point annotations \texttt{linRC(vD,\,b)} take
a value and a Boolean parameters and invoke the non-deterministic abstract
operation and disambiguate it by assigning the parameters to the unspecified
\texttt{vD} and \texttt{b} of the specification. There are two linearisation
points together with the invocations of \texttt{read} (line~6) and
\texttt{write} (line~8). If \texttt{read} fails, then we linearise forcing the
unspecified \texttt{vD} to be \texttt{undef} (line~6), which ensures that the
abstract operation fails without adding any value to \texttt{abs\verbund
  valsRC} nor updating the round \texttt{abs\verbund roundRC}. Otherwise, if
\texttt{write} succeeds with value \texttt{v}, then we linearise forcing the
unspecified value \texttt{vD} and Boolean \texttt{b} to be \texttt{v} and
\texttt{true} respectively (line~8). This ensures that the abstract operation
succeeds and updates the round \texttt{abs\verbund roundRC} to \texttt{k} and
assigns \texttt{v} to the decided value \texttt{abs\verbund vRC}. If
\texttt{write} fails then we linearise forcing the unspecified \texttt{vD} and
\texttt{b} to be \texttt{v} and \texttt{false} respectively (line~9). This
ensures that the abstract operation fails.

\begin{theorem}
  \label{th:round-based-consensus}
  The implementation of Round-Based Consensus in
  Figure~\ref{fig:implementation-paxos-rb-consensus} linearises with respect
  to its specification on the top of
  Figure~\ref{fig:spec-inst-imp-round-based-consensus}.
\end{theorem}

\begin{figure}[t]
  \hspace{1.5em}
  \begin{minipage}[t]{.5\linewidth}
\begin{lstlisting}
read(int k) {
  (@$\langle$@) val vD := random();
    bool b := random(); val v;
    assume(vD (@$\in$@) valsRR);
    assume(pid() =
      ((k - 1) mod (@$n$@)) + 1);
    if (b) {
      if (k >= roundRR) {
        roundRR := k;
        if (!(vRR = undef)) {
          v := vRR; }
        else { v := vD; } }
      else { v := vD; }
      return (true, v); }
    else { return (false, _); } (@$\rangle$@) }
\end{lstlisting}
  \end{minipage}
  \hfill
  \begin{minipage}[t]{.4\linewidth}
\begin{lstlisting}[firstnumber=16]
val vRR := undef;
int roundRR := 0;
set of val valsRR := {undef};

write(int k, val vW) {
  (@$\langle$@) bool b := random();
    assume(!(vW = undef));
    assume(pid() =
      ((k - 1) mod (@$n$@)) + 1);
    valsRR := valsRR (@$\cup$@) {vW};
    if (b && (k >= roundRR)) {
      roundRR := k;
      vRR := vW;
      return true; }
    else { return false; } (@$\rangle$@) }
\end{lstlisting}
  \end{minipage}
  \caption{Specification of \emph{Round-Based Register}.}
  \label{fig:spec-round-based-register}
\end{figure}

\subsubsection{Module \emph{Round-Based Register}.}
Figure~\ref{fig:spec-round-based-register} shows the module's
non-deterministic specification.
Global variable \texttt{vRR} represents the decided value, initially
\texttt{undef}. Global variable \texttt{roundRR} represents the current round,
initially $0$, and global set of values \texttt{valsRR}, initially containing
\texttt{undef}, stores values that may have been proposed by some
proposer. The specification is non-deterministic in that method \texttt{read}
has unspecified local Boolean \texttt{b} and local value \texttt{vD} (we
assume that \texttt{vD} is \texttt{valsRR}), and method \texttt{write} has
unspecified local Boolean \texttt{b}. We assume the current process identifier
is $((\texttt{k}-1)\mmod n)+1$.

Let us explain the specification of the \texttt{read} operation. The operation
can succeed regardless of the proposer's round \texttt{k}, depending on the
value of the unspecified Boolean \texttt{b}. If \texttt{b} is \texttt{true}
and the proposer's round \texttt{k} is valid (line~8), then the read round is
updated to \texttt{k} (line~9) and the operation returns \texttt{(true,\,v)}
(line~14), where \texttt{v} is the read value, which coincides with the
decided value if some decision was committed already or with \texttt{vD}
otherwise. Now to the specification of operation \texttt{write}. The value
\texttt{vW} is always added to the set \texttt{valsRR} (line~25). If the
unspecified Boolean \texttt{b} is false (the round will be stolen by a
posterior operation) or if the round \texttt{k} is non-valid, then the
operation returns \texttt{false} (lines~26 and 30). Otherwise, the current
round is updated to \texttt{k}, and the decided value \texttt{vRR} is updated
to \texttt{vW} and the operation returns \texttt{true} (lines~27--29).

In order to prove that the implementation in
Figures~\ref{fig:implementation-rb-register-read-write} and
\ref{fig:implementation-rb-register-acceptor} linearises with respect to the
specification in Figure~\ref{fig:spec-round-based-register}, we use the
instrumented implementation in Figures~\ref{fig:instrumented-read-write} and
\ref{fig:instrumented-acceptor}, which uses prophecy variables
\cite{Abadi-Lamport:LICS88,Vafeiadis:PhD} that ``guess'' whether the execution
of the method will reach a particular program location or not. The
instrumented implementation also uses external linearisation points. In
particular, the code of the acceptors may help to linearise some of the
invocations to \texttt{read} and \texttt{write}, based on the prophecies and
on auxiliary variables that count the number of acknowledgements sent by
acceptors after each invocation of a \texttt{read} or a \texttt{write}. The
next paragraphs elaborate on our use of prophecy variables and on our helping
mechanism.

\begin{figure}
  \hspace{1.5em}
\begin{minipage}{1.0\linewidth}
\begin{lstlisting}
(@\hl{val abs\verbund vRR := undef; int abs\verbund roundRR := 0;}@)
(@\hl{set of val abs\verbund valsRR := \verblbrace undef\verbrbrace;}@)
(@\hl{single val abs\verbund res\verbund r[1..$\infty$] := undef;}@)
(@\hl{single val abs\verbund res\verbund w[1..$\infty$] := undef;}@)
(@\hl{int count\verbund r[1..$\infty$] := 0; int count\verbund w[1..$\infty$] := 0;}@)
(@\hl{single (bool $\times$ val) proph\verbund r[1..$\infty$] := undef;}@)
(@\hl{single bool proph\verbund w[i..$\infty$] := undef;}@)
read(int k) {
  int j; val v; set of int Q; int maxKW; val maxV; msg m;
  (@\hl{assume(pid() = ((k - 1) mod $n$) + 1);}@)
  (@\hl{$\langle$ if (\textrm{operation reaches} PL: RE\verbund SUCC \textrm{and define} $v = \texttt{maxV}$ \textrm{at that time}) \verblbrace}@)
      (@\hl{proph\verbund r[k] := (true, $v$); \verbrbrace}@)
    (@\hl{else \verblbrace~if (\textrm{operation reaches} PL: RE\verbund FAIL) \verblbrace}@)
              (@\hl{proph\verbund r[k] := (false, \verbund); \verbrbrace~\verbrbrace~$\rangle$}@)
  for (j := 1, j <= (@$n$@), j++) { send(j, [RE, k]); }
  maxKW := 0; maxV := undef; Q := {};
  do { (j, m) := receive();
        switch (m) {
          case [ackRE, @k, v, kW]:
            Q := Q (@$\cup$@) {j};
            if (kW >= maxKW) { maxKW := kW; maxV := v; }
          case [nackRE, @k]:
            (@\hl{$\langle$ linRE(k, undef, false); proph\verbund r[k] := undef; }@)
              return (false, _); (@\hl{$\rangle$ // PL: RE\verbund FAIL}@)
        } if ((@$|\texttt{Q}|$@) = (@$\lceil$@)(n+1)/2(@$\rceil$@)) {
             return (true, maxV); } } (@\hl{// PL: RE\verbund SUCC}@)
  while (true); }
write(int k, val vW) {
  int j; set of int Q; msg m;
  (@\hl{assume(!(vW = undef)); assume(pid() = ((k - 1) mod $n$) + 1);}@)
  (@\hl{$\langle$ if (\textrm{operation reaches} PL: WR\verbund SUCC) \verblbrace~proph\verbund w[k] := true; \verbrbrace}@)
    (@\hl{else \verblbrace~if (\textrm{operation reaches} PL: WR\verbund FAIL) \verblbrace}@)
              (@\hl{proph\verbund w[k] := false; \verbrbrace~\verbrbrace~$\rangle$}@)
  for (j := 1, j <= (@$n$@), j++) { send(j, [WR, k, vW]); }
  Q := {};
  do { (j, m) := receive();
        switch (m) {
          case [ackWR, @k]:
            Q := Q (@$\cup$@) {j};
          case [nackWR, @k]:
            (@\hl{$\langle$ if (count\verbund w[k] = 0) \verblbrace}@)
                (@\hl{linWR(k, vW, false); proph\verbund w[k] := undef; \verbrbrace}@)
              return false; (@\hl{$\rangle$ // PL: WR\verbund FAIL}@)
        } if ((@$|\texttt{Q}|$@) = (@$\lceil$@)(n+1)/2(@$\rceil$@)) {
             return true; } } (@\hl{// PL: WR\verbund SUCC}@)
  while (true); }
\end{lstlisting}
\end{minipage}
  \caption{Instrumented implementation of \texttt{read} and
    \texttt{write} methods.}
  \label{fig:instrumented-read-write}
\end{figure}

Variables \absvRR, \absroundRR\ and \absvalsRR\ in
Figure~\ref{fig:instrumented-read-write} model the abstract state. They are
initially set to \Undefined, $0$ and the set containing \Undefined\
respectively. Variable \absresrb{$k$} is an infinite array of
single-assignment pairs of Boolean and value that model the abstract results
of the invocations to \texttt{read}. (Think of an infinite array as a map from
integers to some type; we use the array notation for convenience.)  Similarly,
variable \absreswb{$k$} is an infinite array of single-assignment Booleans
that models the abstract results of the invocations to \texttt{write}. All the
cells in both arrays are initially \texttt{undef} (\eg\ the initial maps are
empty). Variables \countrb{$k$} and \countwb{$k$} are infinite arrays of
integers that model the number of acknowledgements sent (but not necessarily
received yet) from acceptors in response to respectively read or write
requests. All cells in both arrays are initially $0$. The variable
\prophrb{$k$} is an infinite array of single-assignment pairs
\texttt{bool\,$\times$\,val}, modelling the prophecy for the invocations of
\texttt{read}, and variable \prophwb{$k$} is an infinite array of
single-assignment Booleans modelling the prophecy for the invocations of
\texttt{write}.

The linearisation-point annotations \texttt{linRE(k,\,vD,\,b)} for
\texttt{read} take the proposer's round \texttt{k}, a value \texttt{vD} and a
Boolean \texttt{b}, and they invoke the abstract operation and disambiguate it
by assigning the parameters to the unspecified \texttt{vD} and \texttt{b} of
the specification on the left of
Figure~\ref{fig:spec-round-based-register}. At the beginning of a
\texttt{read(k)} (lines~11--14 of Figure~\ref{fig:instrumented-read-write}),
the prophecy \prophrb{\texttt{k}} is set to \texttt{(true,\,$v$)} if the
invocation reaches \texttt{PL:\,RE\verbund SUCC} in line~26. The $v$ is
defined to coincide with \texttt{maxV} at the time when that location is
reached. That is, $v$ is the value accepted at the greatest round by the
acceptors acknowledging so far, or undefined if no acceptor ever accepted any
value. If the operation reaches \texttt{PL:\,RE\verbund FAIL} in line~24
instead, the prophecy is set to \texttt{(false,\,\verbund)}. (If the method
never returns, the prophecy is left \texttt{undef} since it will never
linearise.) A successful \texttt{read(k)} linearises in the code of the
acceptor in Figure~\ref{fig:instrumented-acceptor}, when the
$\lceil(n+1)/2\rceil$th acceptor sends \texttt{[ackRE,\,k,\,v,\,w]}, and only
if the prophecy is \texttt{(true,\,$v$)} and the operation was not linearised
before (lines~10--14). We force the unspecified \texttt{vD} and \texttt{b} to
be $v$ and \texttt{true} respectively, which ensures that the abstract
operation succeeds and returns \texttt{(true,\,$v$)}. A failing
\texttt{read(k)} linearises at the \texttt{return} in the code of
\texttt{read} (lines~23--24 of Figure~\ref{fig:instrumented-read-write}),
after the reception of \texttt{[nackRE,\,k]} from one acceptor. We force the
unspecified \texttt{vD} and \texttt{b} to be \texttt{undef} and \texttt{false}
respectively, which ensures that the abstract operation fails.

\begin{figure}[t]
  \hspace{1.5em}
\begin{minipage}{1.0\linewidth}
\begin{lstlisting}
process Acceptor(int j) {
  val v := undef; int r := 0; int w := 0;
  start() {
    int i; msg m; int k;
    do { (i, m) := receive();
          switch (m) {
            case [RE, k]:
              if (k < r) { send(i, [nackRE, k]); }
              else { (@$\langle$@) r := k;
                        (@\hl{if (abs\verbund res\verbund r[k] = undef) \verblbrace}@)
                          (@\hl{if (proph\verbund r[k] = (true, $v$)) \verblbrace}@)
                            (@\hl{if (count\verbund r[k] = $\lceil$(n+1)/2$\rceil$ - 1) \verblbrace}@)
                              (@\hl{linRE(k, $v$, true); \verbrbrace~\verbrbrace~\verbrbrace}@)
                        (@\hl{count\verbund r[k]++;}@) send(i, [ackRE, k, v, w]); (@$\rangle$@) }
            case [WR, k, vW]:
              if (k < r) { send(j, i, [nackWR, k]); }
              else { (@$\langle$@) r := k; w := k; v := vW;
                        (@\hl{if (abs\verbund res\verbund w[k] = undef) \verblbrace}@)
                          (@\hl{if (!(proph\verbund w[k] = undef)) \verblbrace}@)
                            (@\hl{if (proph\verbund w[k]) \verblbrace}@)
                              (@\hl{if (count\verbund w[k] = $\lceil$(n+1)/2$\rceil$ - 1) \verblbrace}@)
                                (@\hl{linWR(k, vW, true); \verbrbrace~\verbrbrace}@)
                            (@\hl{else \verblbrace~linWR(k, vW, false); \verbrbrace~\verbrbrace~\verbrbrace}@)
                        (@\hl{count\verbund w[k]++;}@) send(j, i, [ackWR, k]); (@$\rangle$@) }
          } }
    while (true); } }
\end{lstlisting}
\end{minipage}
\caption{Instrumented implementation of acceptor processes.}
\label{fig:instrumented-acceptor}
\end{figure}

The linearisation-point annotations \texttt{linWR(k,\,vW,\,b)} for
\texttt{write} take the proposer's round \texttt{k} and value \texttt{vW}, and
a Boolean \texttt{b}, and they invoke the abstract operation and disambiguate
it by assigning the parameter to the unspecified \texttt{b} of the
specification on the right of Figure~\ref{fig:spec-round-based-register}. At
the beginning of a \texttt{write(k,\,vW)} (lines~31--33 of
Figure~\ref{fig:instrumented-read-write}), the prophecy \prophrb{\texttt{k}}
is set to \texttt{true} if the invocation reaches \texttt{PL:\,WR\verbund
  SUCC} in line~45, or to \texttt{false} if it reaches \texttt{PL:\,WR\verbund
  FAIL} in line~43 (or it is left \texttt{undef} if the method never
returns). A successfully \texttt{write(k,\,vW)} linearises in the code of the
acceptor in Figure~\ref{fig:instrumented-acceptor}, when the
$\lceil(n+1)/2\rceil$th acceptor sends \texttt{[ackWR,\,k]}, and only if the
prophecy is \texttt{true} and the operation was not linearised before
(lines~17--24). We force the unspecified \texttt{b} to be \texttt{true}, which
ensures that the abstract operation succeeds deciding value \texttt{vW} and
updates \texttt{roundRR} to \texttt{k}. A failing \texttt{write(k,\,vW)} may
linearise either at the \texttt{return} in its own code (lines~41--43 of
Figure~\ref{fig:instrumented-read-write}) if the proposer received one
\texttt{[nackWR,\,k]} and no acceptor sent any \texttt{[ackWR,\,k]} yet, or at
the code of the acceptor, when the first acceptor sends \texttt{[ackWR,\,k]},
and only if the prophecy is \texttt{false} and the operation was not
linearised before. In both cases, we force the unspecified \texttt{b} to be
\texttt{false}, which ensures that the abstract operation fails.

\begin{theorem}
  \label{th:round-based-register}
  The implementation of Round-Based Register in
  Figures~\ref{fig:instrumented-read-write} and
  \ref{fig:instrumented-acceptor} linearises with respect to its specification
  in Figure~\ref{fig:spec-round-based-register}.
\end{theorem}



\section{Multi-Paxos via Network Transformations}
\label{sec:multi-paxos-via}

We now turn to more complicated distributed protocols that build upon the idea
of Paxos consensus. Our ultimate goal is to reuse the verification result from
the Sections~\ref{sec:faithfull-deconstruction}--\ref{sec:nodet-specs}, as
well as the high-level round-based register interface.  In this section, we
will demonstrate how to reason about an implementation of \mpaxos as of an
array of \emph{independent} instances of the \emph{Paxos} module defined
previously, despite the subtle dependencies between its sub-components, as
present in \mpaxos's ``canonical''
implementations~\cite{VanRenesse-Altinbuken:ACS15,Lamport:TOPLAS98,Chand-al:FM16}.
While an abstraction of \mpaxos to an array of independent shared
``single-shot'' registers is almost folklore, what appears to be inherently
difficult is to verify a \mpaxos-based consensus (\wrt to the array-based
abstraction) by means of \emph{reusing} the proof of a SD-Paxos. All proofs of
\mpaxos we are aware of are, thus, \emph{non-modular} with respect to
underlying SD-Paxos
instances~\cite{Padon-al:OOPSLA17,Chand-al:FM16,Rahli-al:AVOCS15}, \ie, they
require one to redesign the invariants of the \emph{entire} consensus
protocol.


This proof modularity challenge stems from the optimised nature of a classical
\mpaxos protocol, as well as its real-world
implementations~\cite{Chandra-al:PODC07}.
In this part of our work is to distil such protocol-aware optimisations into
a separate \emph{network semantics layer}, and show that each of them refines
the semantics of a Cartesian product-based view, \ie, exhibits the very same
client-observable behaviours.
%
%
To do so, we will establishing the refinement between the optimised
implementations of \mpaxos and a simple Cartesian product abstraction, which
will allow to extend the register-based abstraction, explored before in this
paper, to what is considered to be a canonical amortised \mpaxos
implementation.



\subsection{Abstract Distributed Protocols}
\label{sec:comb-def}

We start by presenting the formal definitions of encoding distributed
protocols (including Paxos), their message vocabularies, protocol-based
network semantics, and the notion of an observable~behaviours.

%


\paragraph{Protocols and messages.}
\label{sec:prot-mess-vocab}

\begin{wrapfigure}[8]{r}[0pt]{0.49\textwidth} 
\centering 
\vspace{-25pt}
\begin{minipage}[l]{4.6cm}
{\small{
\[
\!\!\!\!\!\!\!\!\!\!
\begin{array}{l@{\ \ }r@{\ }c@{\ }l}
\text{Protocols} & \Protocol \ni p & \eqdef & \angled{\Lstate, \mvoc,
                                              \step} \\
\text{Configurations} & \Gstate \ni \gstate & \eqdef & \Nodes
                                                            \pfun \Lstate \\
\text{Internal steps} & \stepi & \in & \Lstate \times \Lstate
\\
\text{Receive-steps} & \stepr & \in & \Lstate \times \mvoc \times \Lstate
\\
\text{Send-steps} & \steps & \in & \Lstate \times \Lstate \times \powerset{\mvoc}
\end{array}
\]
}}
\end{minipage}
\caption{States and transitions.}
\label{fig:basic}
\end{wrapfigure}
Figure~\ref{fig:basic} provides basic definitions of the distributed protocols
and their components.
Each protocol $p$ is a tuple
$\angled{\Lstate, \mvoc, \stepi, \stepr, \steps}$.  $\Lstate$ is a set of
local states, which can be assigned to each of the participating nodes, also
determining the node's role via an additional tag,\footnote{We leave out
  implicit the consistency laws for the state, that are protocol-specific.} if
necessary (\eg, an acceptor and a proposer states in Paxos are different).
%
%
$\mvoc$ is a ``message vocabulary'', determining the set of messages that can
be used for communication between the nodes.

Messages can be thought of as JavaScript-like dictionaries, pairing unique
fields (isomorphic to strings) with their values. For the sake of a uniform
treatment, we assume that each message $m \in \mvoc$ has at least two fields,
$\From$ and $\To$ that point to the source and the destination node of a
message, correspondingly. In addition to that, for simplicity we will assume
that each message carries a Boolean field $\Active$, which is set to $\True$
when the message is sent and is set to $\False$ when the message is received
by its destination node. This flag is required to keep history information
about messages sent in the past, which is customary in frameworks for
reasoning about distributed
protocols~\cite{Padon-al:PLDI16,Hawblitzel-al:SOSP15,Wilcox-al:SNAPL17}. 
%
%
We assume that a ``message soup'' $M$ is a multiset of messages (\ie\, a set
with zero or more copies of each message) and we consider that each copy of
the same message in the multiset has its own ``identity'', and we write
$m \neq m'$ to represent that $m$ and $m'$ are not the same copy of a
particular message. 
%

Finally, $\step_{\set{\text{int}, \text{rcv}, \text{snd}}}$ are step-relations
that correspond to the internal changes in the local state of a node
($\stepi$), as well as changes associated with sending ($\steps$) and
receiving ($\stepr$) messages by a node, as allowed by the protocol.
Specifically, $\stepi$ relates a local node state before and after the allowed
internal change; $\stepr$ relates the initial state and an incoming message
$m \in \mvoc$ with the resulting state; $\steps$ relates the internal state,
the output state and the set of atomically sent messages. For simplicity we
will assume that $\id \subseteq \stepi$.

In addition, we consider $\Lstate_0 \subseteq \Lstate$---the set of the
allowed \emph{initial} states, in which the system can be present at the very
beginning of its execution.
The global state of the network $\gstate \in \Gstate$ is a map from node
identifiers ($n \in \Nodes$) to local states from the set of states $\Lstate$,
defined by the protocol.

\paragraph{Simple network semantics.}
\label{sec:network-semantics}

\begin{figure}[t]
\centering
\begin{mathpar}
\inferrule*[Lab=StepInt,width=5cm]
{
n \in \dom{\gstate}
\\
\lstate = \gstate(n)
\\
\angled{\lstate, \lstate'} \in p.\stepi 
\\
\gstate' = \gstate[n \mapsto \lstate']   
}
{\angled{\gstate, M} \opstepi{p} \angled{\gstate', M}}
\and
\inferrule*[Lab=StepSend,width=7cm]
{
n \in \dom{\gstate}
\\
\lstate = \gstate(n)
\\
\angled{\lstate, \lstate', \ms} \in p.\steps 
\\
\gstate' = \gstate[n \mapsto \lstate']
\\
M' = M \cup \ms
}
{\angled{\gstate, M} \opsteps{p} \angled{\gstate', M'}}
\and
\inferrule[StepReceive]
{
m \in M
\\
m.\Active
\\
m.\To \in \dom{\gstate}
\\
\lstate = \gstate(m.\To)
\\
\angled{\lstate, m, \lstate'} \in p.\stepr 
\\
m' = m[\Active \mapsto \False]
\\
\gstate' = \gstate[n \mapsto \lstate']
\\
M' = M \setminus \set{m} \cup \set{m'}
}
{\angled{\gstate, M} \opstepr{p} \angled{\gstate', M'}}
\end{mathpar}  
\caption{Transition rules of the simple protocol-aware network semantics}
\label{fig:sem}
\end{figure}

The simple initial operational semantics of the network
($\opstep{p}~\subseteq~(\Gstate \times \powerset{\mvoc}) \times (\Gstate
\times \powerset{\mvoc})$)
is parametrised by a protocol $p$ and relates the initial \emph{configuration}
(\ie, the global state and the set of messages) with the resulting
configuration. It is defined via as a reflexive closure of the union of three
relations $\opstepi{p} \cup \opstepr{p} \cup \opsteps{p}$, their rules are
given in Figure~\ref{fig:sem}.

The rule~\rulename{StepInt} corresponds to a node $n$ picked
non-deterministically from the domain of a global state $\gstate$, executing
an internal transition, thus changing its local state from $\lstate$
to~$\lstate'$.
The rule~\rulename{StepReceive} non-deterministically picks a $m$ message from
a message soup $M \subseteq \mvoc$, changes the state using the protocol's
receive-step relation $p.\stepr$ at the corresponding host node $\To$, and
updates its local state accordingly in the common mapping
($\gstate[\To \mapsto \lstate']$).
Finally, the rule~\rulename{StepSend}, non-deterministically picks a node $n$,
executes a send-step, which results in updating its local state emission of a
set of messages $\ms$, which is added to the resulting soup.
In order to ``bootstrap'' the execution, the initial states from the set
$\Lstate_0 \subseteq \Lstate$ are assigned to the nodes.

We next define the observable protocol behaviours \wrt the simple network
semantics as the prefix-closed set of all system's configuration traces.
\begin{definition}[Protocol behaviours]
\label{def:beh}
{\small{
\[
\beh{p} = \bigcup_{m \in \Nat} \set{ \angled{\angled{\gstate_0, M_0},
    \ldots, \angled{\gstate_m, M_m}} \left|
    \begin{array}{lll}
      \exists \lstate_0^{n \in N} \in \Lstate_0, &
      \gstate_0 = \biguplus_{n \in N}[n \mapsto
      \lstate_0^n]~\aand \\[3pt]
&                   \angled{\gstate_0, M_0} \opstep{p} \ldots \opstep{p} \angled{\gstate_m, M_m} 
    \end{array}
\right.
}
\]
}}
\end{definition}
That is, the set of behaviours captures all possible configurations of initial
states for a fixed set of nodes $N \subseteq \Nodes$.  In this case, the set
of nodes $N$ is an implicit parameter of the definition, which we fix in the
remainder of this section.

\begin{example}[Encoding SD-Paxos]
\label{ex:simple-paxos}
An abstract distributed protocol for SD-Paxos can be extracted from the
pseudo-code of Section~\ref{sec:faithfull-deconstruction} by providing a
suitable small-step operational semantics \`{a} la Winskel~\cite{Win93}. We
restraint ourselves from giving such formal semantics, but in
\ifdefined\extflag
Appendix~\ref{ap:encoding-SD-Paxos}
\else
Appendix~D of the extended version of the paper
\fi 
we outline how the distributed protocol
would be obtained from the given operational semantics and from the code in
Figures~\ref{fig:implementation-rb-register-read-write},
\ref{fig:implementation-rb-register-acceptor} and
\ref{fig:implementation-paxos-rb-consensus}.

\end{example}

\subsection{Out-of-Thin-Air Semantics.}
\label{sec:out-thin-air}

We now introduce an intermediate version of a simple protocol-aware semantics
that generates messages ``out of thin air'' according to a certain predicate
$\predota \subseteq \Lstate \times \mvoc$, which determines whether the
network generates a certain message without exercising the corresponding
send-transition. The rule is as follows:
\begin{mathpar}  
{\small{
\inferrule*[Lab=OTASend,width=10cm]
{
n \in \dom{\gstate}
\\
\lstate = \gstate(n)
\\
\predota(\lstate, m)
\\
M' = M \cup \set{m}
}
{\angled{\gstate, M} \opstepota{p,\predota} \angled{\gstate, M'}}%
}}
\end{mathpar}  
That is, a random message $m$ can be sent at any moment in the
semantics described by $\opstep{p} \cup \opstepota{p,\predota}$, given
that the node $n$, ``on behalf of which'' the message is sent is in a
state $\lstate$, such that $\predota(\lstate, m)$ holds.

\begin{example}
\label{ex:sdpota}
  In the context of Single-Decree Paxos, we can define $\predota$ as follows:
\[
\predota(\lstate, m) \eqdef m.\mcont = \texttt{[RE,\,$k$]} \aand \lstate.\pid = n \aand
\lstate.\role = \mathit{Proposer}  \aand k \leq \lstate.\texttt{kP}
\]
In other words, if a node $n$ is a \emph{Proposer} currently operating with a
round $\lstate.\texttt{kP}$, the network semantics can always send another
request ``on its behalf'', thus generating the message ``out-of-thin-air''.
Importantly, the last conjunct in the definition of $\predota$ is in terms of
$\leq$, rather than equality. This means that the predicate is intentionally
loose, allowing for sending even ``stale'' messages, with expired rounds that
are smaller than what $n$ currently holds (no harm in that!).
\end{example}

By definition of single-decree Paxos protocol, the following lemma
holds:

\begin{lemma}[OTA refinement]
  \label{lm:otaref}
  $\beh{\opstep{p} \cup \opstepota{p,\predota}} \subseteq \beh{{p}}$, where
  $p$ is an instance of the module Paxos, as defined in
  Section~\ref{sec:faithfull-deconstruction} and in
  Example~\ref{ex:simple-paxos}.
\end{lemma}


\subsection{Slot-Replicating Network Semantics.}
\label{sec:basic-comb-their}

With the basic definitions at hand, we now proceed to describing alternative
network behaviours that make use of a specific protocol
$p = \angled{\Lstate, \mvoc, \stepi, \stepr, \steps}$, which we will consider
to be fixed for the remainder of this section, so we will be at times
referring to its components (\eg, $\stepi$, $\stepr$, \etc) without a
qualifier.

\begin{figure}[t]
\centering
{\small{
\begin{mathpar}
\inferrule*[Lab=SRStepInt, width=5cm]
{
i \in I
\\
n \in \dom{\gstate}
\\
\lstate = \gstate(n)[i]
\\
\angled{\lstate, \lstate'} \in p.\stepi 
\\
\gstate' = \gstate[n[i] \mapsto \lstate']
}
{\angled{\gstate, M} \opstepi{\cartp} \angled{\gstate', M}}
\and
\inferrule*[Lab=SRStepSend,width=7cm]
{
i \in I 
\\
n \in \dom{\gstate}
\\
\lstate = \gstate(n)[i]
\\
\angled{\lstate, \lstate', \ms} \in p.\steps 
\\
\gstate' = \gstate[n[i] \mapsto \lstate']
\\
M' = M \cup \ms[\Slot \mapsto i]
}
{\angled{\gstate, M} \opsteps{\cartp} \angled{\gstate', M'}}
\and
\inferrule[SRStepReceive]
{
m \in M
\quad
m.\Active
\\
m.\To \in \dom{\gstate}
\\
\lstate = \gstate(m.\To)[m.\Slot]
\\
\angled{\lstate, m, \lstate'} \in p.\stepr 
\\
m' = m[\Active \mapsto \False]
\\
\gstate' = \gstate(n)[m.\Slot \mapsto \lstate']
\\
M' = M \setminus \set{m} \cup \set{m'}
}
{\angled{\gstate, M} \opstepr{\cartp} \angled{\gstate', M'}}
\end{mathpar}  
}}
\caption{Transition rules of the slot-replicating network semantics.}
\label{fig:cart-sem}
\end{figure}

Figure~\ref{fig:cart-sem} describes a semantics of a \emph{slot-replicating}
(SR) network that exercises multiple copies of the \emph{same} protocol
instance $p_i$ for $i \in I$, some, possibly infinite, set of indices, to
which we will be also referring as \emph{slots}. Multiple copies of the
protocol are incorporated by enhancing the messages from $p$'s vocabulary
$\mvoc$ with the corresponding indices, and implementing the on-site dispatch
of the indexed messages to corresponding protocol instances at each node. The
local protocol state of each node is, thus, no longer a single element being
updated, but rather an \emph{array}, mapping $i \in I$ into $\lstate_i$---the
corresponding local state component.
The small-step relation for SR semantics is denoted by $\opstep{\cartp}$. The
rule \rulename{SRStepInt} is similar to \rulename{StepInt} of the simple
semantics, with the difference that it picks not only a node but also an index
$i$, thus referring to a specific component $\gstate(n)[i]$ as $\lstate$ and
updating it correspondingly $(\gstate(n)[i] \mapsto \lstate')$.
For the remaining transitions, we postulate that the messages from $p$'s
vocabulary $p.\mvoc$ are enhanced to have a dedicated field $\Slot$, which
indicates a protocol copy at a node, to which the message is directed. 
The receive-rule~\rulename{SRStepReceive} is similar
to~\rulename{StepReceive} but takes into the account the value of
$m.\Slot$ in the received message $m$, thus redirecting it to the
corresponding protocol instance and updating the local state
appropriately. Finally, the rule~\rulename{SRStepSend} can be now
executed for any slot $i \in I$, reusing most of the logic of the
initial protocol and otherwise mimicking its simple network semantic
counterpart~\rulename{StepSend}.

Importantly, in this semantics, for two different slots $i, j$, such that
$i \neq j$, the corresponding ``projections'' of the state behave
\emph{independently} from each other. Therefore, transitions and messages in
the protocol instances indexed by $i$ at different nodes \emph{do not
  interfere} with those indexed by~$j$.
This observation can be stated formally. In order to do so we first defined
the behaviours of slot-replicating networks and their projections as follows:
\begin{definition}[Slot-replicating protocol behaviours]
\label{def:cart-beh}
{\footnotesize{
\[
\beh{\cartp} = \bigcup_{m \in \Nat} \set{ \angled{\angled{\gstate_0, M_0},
    \ldots, \angled{\gstate_m, M_m}} \left|
    \begin{array}{lll}
      \exists \lstate_0^{n \in N} \in \Lstate_0, \\ 
      \gstate_0 = \biguplus_{n \in N}[n\mapsto
      \set{i \mapsto \lstate_0^n ~|~ i \in I}] & \aand \\
      \angled{\gstate_0, M_0} \opstep{p} \ldots \opstep{p} \angled{\gstate_m, M_m} 
    \end{array}
\right.
}
\]
}}
\end{definition}
That is, the slot-replicated behaviours are merely behaviours with respect to
networks, whose nodes hold \emph{multiple instances} of the same protocol,
indexed by slots $i \in I$.
For a slot $i \in I$, we define \emph{projection} $\beh{\cartp}|_{i}$ as a set
of global state traces, where each node's local states is restricted only to
its $i$th component. The following simulation lemma holds naturally,
connecting the state-replicating network semantics and simple network
semantics.

\begin{lemma}[Slot-replicating simulation]
  \label{lm:sem-sim1} 
  For all $I, i \in I$, $\beh{{\cartp}}|_{i} = \beh{{p}}$.
\end{lemma}

\begin{example}[Slot-replicating semantics and Paxos]
\label{ex:sr-paxos}
  Given our representation of Paxos using roles (acceptors/proposers) encoded
  via the corresponding parts of the local state $\lstate$, we can construct a
  ``na\"{i}ve'' version of \mpaxos by using the SR semantics for the
  protocol. In such, every slot will correspond to a SD Paxos instance, not
  interacting with any other slots. From the practical perspective, such an
  implementation is rather non-optimal, as it does not exploit dependencies
  between rounds accepted at different~slots.
\end{example}

\subsection{Widening Network Semantics.}
\label{sec:widen-netw-semant}

We next consider a version of the SR semantics, extended with a new rule for
handling received messages. In the new semantics, dubbed \emph{widening}, a
node, upon receiving a message $m \in T$, where $T \subseteq p.\mvoc$, for a
slot $i$, \emph{replicates} it for all slots from the index set $I$, for the
very same node. The new rule is as follows:
\begin{mathpar}  
{\small{
\inferrule*[Lab=WStepReceiveT,width=11cm]
{
m \in M
\\
m.\Active
\\
m.\To \in \dom{\gstate}
\\
\lstate = \gstate(m.\To)[m.\Slot]
\\
\angled{\lstate, m, \lstate'} \in p.\stepr 
\\
m' = m[\Active \mapsto \False]
\\
\gstate' = \gstate(n)[m.\Slot \mapsto \lstate']
\\
\ms = \text{if}~(m \in T)~\text{then}~\set{m'~|~m' = m[\Slot \mapsto j], j \in I}~\text{else}~\emptyset
}
{\angled{\gstate, M} \opstepr{\cartw} \angled{\gstate', (M \setminus \set{m}) \cup \set{m'} \cup \ms}}
}}
\end{mathpar}  
At first, this semantics seems rather unreasonable: it might create more
messages than the system can ``consume''. However, it is possible to prove
that, under certain conditions on the protocol $p$, the set of behaviours
observed under this semantics (\ie, with \rulename{SRStepReceive} replaced by
\rulename{WStepReceiveT}) is \emph{not larger} than $\beh{{\cartp}}$ as given
by Definition~\ref{def:cart-beh}.
To state this formally we first relate the set of ``triggering'' messages $T$
from~\rulename{WStepReceiveT} to a specific predicate~$\predota$.

\begin{definition}[OTA-compliant message sets]
\label{def:otacomp}
The set of messages $T \subseteq p.\mvoc$ is OTA-compliant with the predicate
$\predota$ iff  for any $b \in \beh{p}$ and $\angled{\gstate, M} \in b$, if $m \in M$,
  then $\predota(\gstate(m.\from), m)$.
\end{definition}
In other words, the protocol $p$ is relaxed enough to ``justify'' the presence
of $m$ in the soup at \emph{any} execution, by providing the predicate
$\predota$, relating the message to the corresponding sender's state.
Next, we use this definition to slot-replicating and widening semantics via
the following definition.
\begin{definition}[$\predota$-monotone protocols]
\label{def:pota-mono}
  A protocol $p$ is $\predota$-monotone iff for any, $b \in \beh{\cartp}$,
  $\angled{\gstate, M} \in b$, $m$, $i = m.\Slot$, and $j \neq i$,
  if $\predota(\gstate(m.\from)[i], \canon{m})$ then we have that
  $\predota(\gstate(m.\from)[j], \canon{m})$, where $\canon{m}$ ``removes''
  the $\Slot$ field from $m$.
\end{definition}
Less formally, Definition~\ref{def:pota-mono} ensures that in a slot-replicated
product $\cartp$ of a protocol $p$, different components cannot perform ``out of
sync'' \wrt $\predota$. Specifically, if a node in $i$th projection is related to a certain
message $\canon{m}$ via $\predota$, then any other projection $j$ of the same
node will be $\predota$-related to this message, as well.

\begin{example}
  This is a ``non-example''. A version of slot-replicated SD-Paxos, where we
  allow for arbitrary increments of the round \emph{per-slot} at a same
  proposer node (\ie, out of sync), would not be monotone \wrt $\predota$ from
  Example~\ref{ex:sdpota}.
  In contrast, a slot-replicated product of SD-Paxos instances with
  fixed rounds is monotone \wrt the same~$\predota$.
\end{example}



\begin{lemma}
\label{lm:wsem}
If $T$ from \rulename{WStepReceiveT} is OTA-compliant with predicate
$\predota$, such that
$\beh{\opstep{p} \cup \opstepota{p,\predota}} \subseteq
\beh{\opstep{p}}$ and $p$ is $\predota$-monotone,
then $\beh{\opstep{\cartw}} \subseteq \beh{\opstep{\cartp}}$.
\end{lemma}

\begin{example}[Widening semantics and Paxos]
\label{ex:wid-paxos}
The SD-Paxos instance as described in
Section~\ref{sec:faithfull-deconstruction} satisfies the refinement condition
from Lemma~\ref{lm:wsem}. %
By taking $T = \set{m ~|~ m = \set{\mcont = \texttt{[RE,\,k]}; \ldots}}$ %
and using Lemma~\ref{lm:wsem}, we obtain the refinement between widened
semantics and SR semantics of Paxos.
\end{example}

\subsection{Optimised Widening Semantics.}
\label{sec:optim-widen-semant}

Our next step towards a realistic implementation of \mpaxos out of SD-Paxos
instances is enabled by an observation that in the widening semantics, the
replicated messages are \emph{always} targeting the same node, to which the
initial message $m \in T$ was addressed. This means that we can optimise the
receive-step, making it possible to execute multiple receive-transitions of
the core protocol in batch. The following rule~\rulename{OWStepReceiveT}
captures this intuition formally:
\begin{mathpar}
{\small{
\inferrule*[Lab=OWStepReceiveT,width=13cm]
{
m \in M
\\
m.\Active
\\
m.\To \in \dom{\gstate}
\\
\angled{\gstate', \ms} = \reproc(\gstate, n, m)
}
{\angled{\gstate, M} \opstepr{\cartow} \angled{\gstate', M \setminus \set{m}
    \cup \set{m[\Active \mapsto \False]} \cup \ms}}
}}
\end{mathpar}\vspace{-13pt}
\[
{\small{
\begin{array}{l}
\text{where}~\reproc(\gstate, n, m) \eqdef \angled{\gstate', \ms},
  ~\text{such that}~ \ms = \bigcup_{j}\set{m[\Slot \mapsto j]~|~m \in \ms_j},\\[2pt]
  \forall j \in I, \lstate = \gstate(m.\To)[j] \aand
  \angled{\lstate_j, \canon{m}, \lstate_j^{1}}
  \in p.\stepr ~\aand
  \angled{\lstate_j^1, \lstate_j^2} \in p.\stepi^{*} \aand
  \angled{\lstate_j^2, \lstate_j^3,\ms_j} \in p.\steps, \\[2pt]
  \forall j \in I, \gstate'(m.\To)[j] = \lstate_j^3.
\end{array}
}}
\]\smallskip

\noindent
In essence, the rule~$\rulename{OWStepReceiveT}$ blends several steps
of the widening semantics together for a single message: (a) it first
receives the message and replicates it for all slots at a destination
node; (b) performs receive-steps for the message's replicas at each
slot; (c) takes a number of internal steps, allowed by the protocol's
$\stepi$; and (d) takes a send-transition, eventually sending all
emitted message, instrumented with the corresponding slots.

\begin{example}
\label{ex:opt-paxos}
Continuing Example~\ref{ex:wid-paxos}, with the same parameters, the
optimising semantics will execute the transitions of an acceptor, \emph{for
  all slots}, triggered by receiving a single \texttt{[RE,\,k]} message for a
particular slot, sending back \emph{all} the results for all the slots, which
might either agree to accept the value or reject it.
\end{example}

The following lemma relates the optimising and the widening semantics.

\begin{lemma}[Refinement for OW semantics]
\label{lm:row}
For any $b \in \beh{\opstep{\cartow}}$ there exists
$b' \in \beh{\opstep{\cartw}}$, such that $b$ can be obtained from $b'$ by
replacing sequences of configurations
$[\angled{\gstate_k, M_k}, \ldots, \angled{\gstate_{k+m}, M_{k+m}}]$ that have
just a single node $n$, whose local state is affected in
$\gstate_k, \ldots, \gstate_{k+m}$, by
$[\angled{\gstate_k, M_k}, \angled{\gstate_{k+m}, M_{k+m}}]$.
\end{lemma}
That is, behaviours in the optimised semantics are the same as in the
widening semantics, modulo some sequences of locally taken steps that
are being ``compressed'' to just the initial and the final
configurations.

\subsection{Bunching Semantics.}
\label{sec:bunching-semantics}

\begin{figure}[t]
\centering
\centering
\begin{mathpar}
\inferrule*[Lab=BStepRecvB,width=5cm]
{
m \in M
\quad
m.\Active
\quad
m.\To \in \dom{\gstate}
\\
\angled{\gstate', \ms} = \reproc(\gstate, n, m)
\\
M' = M \setminus \set{m}
    \cup \set{m[\Active \mapsto \False]} 
\\
m' = \bunch{\ms, m.\To, m.\From}
}
{\angled{\gstate, M} \opstepr{B} \angled{\gstate', M'\cup
    \set{m'}}}
\and
\inferrule*[Lab=BStepRecvU,width=5cm]
{
m \in M
\quad
m.\Active
\quad
m.\To \in \dom{\gstate}
\\
m.\msgs = \ms
\quad
M' = M \setminus \set{m} \cup \ms} 
{\angled{\gstate, M} \opstepr{B} \angled{\gstate, M'}}%
\end{mathpar}    
\[
{\small{
\text{where}~\bunch{\ms, n_1, n_2} = \set{\msgs = \ms; \From = n_1; \To = n_2;
  \Active = \True}.
}}
\]
\caption{Added rules of the Bunching Semantics}
\label{fig:bunch}
\end{figure}

As the last step towards \mpaxos, we introduce the final network
semantics that optimises executions according to $\opstep{\cartow}$
described in previous section even further by making a simple addition
to the message vocabulary of a slot-replicated SD
Paxos---\emph{bunched messages}.
A bunched message simply packages together several messages, obtained
typically as a result of a ``compressed'' execution via the optimised
semantics from Section~\ref{sec:optim-widen-semant}. We define two new rules
for packaging and ``unpackaging'' certain messages in Figure~\ref{fig:bunch}.
The two new rules can be added to enhance either of the versions of
the slot-replicating semantics shown before. In essence, the only
effect they have is to combine the messages resulting in the execution
of the corresponding steps of an optimised widening (via
$\rulename{BStepRecvB}$), and to unpackage the messages $\ms$ from a
bunching message, adding them back to the soup
($\rulename{BStepRecvU}$).
The following natural refinement result holds:
\begin{lemma}
\label{lm:bunch}
  For any $b \in \beh{\opstep{B}}$ there exists
  $b' \in \beh{\opstep{\cartow}}$, such that $b'$ can be obtained from
  $b$ by replacing all bunched messages in $b$ by their
  $\msgs$-component.
\end{lemma}
The rule $\rulename{BStepRecvU}$ enables effective local caching of the
bunched messages, so they are processed \emph{on demand} on the recipient side
(\ie, by the per-slot proposers), allowing the implementation to \emph{skip}
an entire round of Phase~1.

\subsection{The Big Picture.}
\label{sec:big-picture}

\begin{figure}[t]
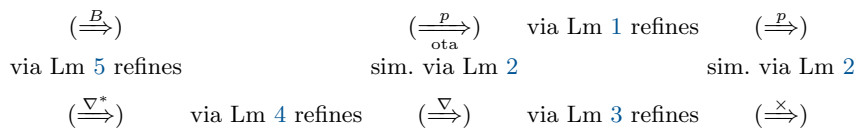

  \centering
{\small{
\[
\begin{array}{c@{\ }c@{\ }c@{\ }c@{\ }c}
(\opstep{B})  &&
(\opstepota{p}) & \text{via Lm~\ref{lm:otaref}~refines} & 
(\opstep{p})
\\[4pt]
 \text{via Lm~\ref{lm:bunch}~refines} &&
\text{sim. via~Lm~\ref{lm:sem-sim1}} && 
\text{sim. via~Lm~\ref{lm:sem-sim1}}
\\[4pt]
(\opstep{\cartow}) & \text{via Lm~\ref{lm:row}~refines} & 
(\opstep{\cartw}) & \text{via Lm~\ref{lm:wsem}~refines} &
(\opstep{\cartp})
\end{array}
\]  
}}
\caption{Refinement between different network semantics.}
\label{fig:ref}
\end{figure}

What exactly have we achieved by introducing the described above family of
semantics? As illustrated in Figure~\ref{fig:ref}, all behaviours of the
leftmost-topmost, bunching semantics, which corresponds precisely to an implementation
of \mpaxos with an ``amortised'' Phase~1, can be transitively related to the
corresponding behaviours in the rightmost, vanilla slot-replicated version of
a simple semantics (via the correspondence from Lemma~\ref{lm:otaref}) by
constructing the corresponding refinement
mappings~\cite{Abadi-Lamport:LICS88}, delivered by the proofs of
Lemmas~\ref{lm:wsem}--\ref{lm:bunch}.

From the perspective of Rely/Guarantee reasoning, which was employed in
Section~\ref{sec:nodet-specs}, the refinement result from Figure~\ref{fig:ref}
justifies the replacement of a semantics on the right of the diagram by one to
the left of it, as all program-level assertions will remain substantiated by
the corresponding system configurations, as long as they are \emph{stable}
(\ie, resilient \wrt transitions taken by nodes different from the one being
verified), which they are in our case.



\section{Putting It All Together}
\label{sec:putting-it-all}

We culminate our story of faithfully deconstructing and abstracting
Paxos via a round-based register, as well as recasting \mpaxos via a
series of network transformations, by showing how to \emph{implement}
the register-based abstraction from
Section~\ref{sec:faithfull-deconstruction} in tandem with the network
semantics from Section~\ref{sec:multi-paxos-via} in order to deliver
provably correct, yet efficient, implementation of \mpaxos.

\begin{figure}[t]
\hspace{1.5em}
\begin{minipage}{.5\linewidth}
\begin{lstlisting}
proposeM(val^ v, val v0) {
  (@$\langle$@) assume(!(v0 = undef));
    if (*v = undef) { *v := v0; }
    return *v; (@$\rangle$@) }
\end{lstlisting}
\end{minipage}
\begin{minipage}{.41\linewidth}
\begin{lstlisting}[firstnumber=5]
val vM[1..(@$\infty$@)] := undef;
getR(int s) { return &(vM[s]); }
proposeM(getR(1), v);
proposeM(getR(2), v);
\end{lstlisting}
\end{minipage}
\caption{Specification of \emph{\mpaxos} and interaction via a \emph{register
    provider}. }
  \label{fig:spec-multi-paxos}
\end{figure}

The crux of the composition of the two results---a register-based abstraction
of SD Paxos and a family of semantics-preserving network transformations---is
a convenient interface for the end client, so she could interact with a
consensus instance via the \texttt{proposeM} method in lines~1--4 of
Figure~\ref{fig:spec-multi-paxos}, no matter with which particular slot of a
\mpaxos implementation she is interacting.
To do so, we propose to introduce a \emph{register provider}---a service that
would give a client a ``reference'' to the consensus object to interact
with. Lines~6--7 of Figure~\ref{fig:spec-multi-paxos} illustrate the
interaction with the service provider, where the client requests two specific
slots, 1 and 2, of \mpaxos by invoking \texttt{getR} and providing a slot
parameter. In both cases the client proposes the very same value \texttt{v} in
the two instances that run the same machinery. (Notice that, except for the
reference to the consensus object, \texttt{proposeM} is identical to the
\texttt{proposeP} on the right of Figure~\ref{fig:spec-paxos}, which we have
verified \wrt linearisability in Section~\ref{sec:faithfull-deconstruction}.)

The implementation of \mpaxos that we have in mind resembles the one in
Figures~\ref{fig:implementation-rb-register-read-write},
\ref{fig:implementation-rb-register-acceptor} and
\ref{fig:implementation-paxos-rb-consensus} of
Section~\ref{sec:faithfull-deconstruction}, but where all the global data is
provided by the register provider and passed by reference. What differs in
this implementation with respect to the one in
Section~\ref{sec:faithfull-deconstruction} and is hidden from the client is
the semantics of the network layer used by the bottom layer (\cf~ left part of
Figure~\ref{fig:spec-paxos}) of the register-based implementation. The \mpaxos
instances run (without changing the register's code) over this network layer,
which ``overloads'' the meaning of the \texttt{send/receive} primitives from
Figures~\ref{fig:implementation-rb-register-read-write}
and~\ref{fig:implementation-rb-register-acceptor} to follow the bunching
network semantics, described in Section~\ref{sec:bunching-semantics}.

\begin{theorem}
  \label{thm:grand-theorem}
  The implementation of \mpaxos that uses a register provider and bunching
  network semantics refines the specification in
  Figure~\ref{fig:spec-multi-paxos}.
\end{theorem}

We implemented the register/network semantics in a proof-of-concept prototype
written in Scala/Akka.\footnote{The code is available
  at~\url{https://github.com/certichain/protocol-combinators}.}
We relied on the abstraction mechanisms of Scala, allowing us to implement the
register logic, verified in Section~\ref{sec:nodet-specs}, separately from the
network middle-ware, which has provided a family of Semantics from
Section~\ref{sec:multi-paxos-via}.
Together, they provide a family of provably correct, modularly verified
\emph{distributed} implementations, coming with a simple \emph{shared
  memory-like} interface.



\section{Related Work}
\label{sec:related-work}


\paragraph{Proofs of Linearisability via Rely/Guarantee.}

Our work builds on the results of Boichat~\etal~\cite{Boichat-al:SN03}, who
were first to propose to a systematic deconstruction of Paxos into
read/write operations of a \emph{round-based register} abstraction.
We extend and harness those abstractions, by intentionally introducing
more non-determinism into them, which allows us to provide the first
modular (\ie, mutually independent) proofs of Proposer and Acceptor
using Rely/Guarantee with linearisation points and prophecies.
While several logics have been proposed recently to prove linearisability of
concurrent implementations using Rely/Guarantee
reasoning~\cite{Liang-Feng:PLDI13,Vafeiadis:PhD,Liang-Feng:POPL16,Khyzha-al:FM16},
none of them considers message-passing distributed systems or consensus
protocols.

\paragraph{Verification of Paxos-family Algorithms.}

Formal verification of different versions of Paxos-family protocols \wrt
inductive invariants and liveness has been a focus of multiple verification
efforts in the past fifteen years.
To name just a few, Lamport has specified and verified Fast
Paxos~\cite{Lamport:DC06} using TLA+ and its accompanying model
checker~\cite{Yu-al:CHARME99}. 
Chand~\etal used TLA+ to specify and verify \mpaxos implementation,
similar to the one we considered in this work~\cite{Chand-al:FM16}.
A version of SD-Paxos has been verified by Kellomaki using
the PVS theorem prover~\cite{Kellomaki:04}.
Jaskelioff and Merz have verified Disk Paxos in
Isabelle/HOL~\cite{Jaskelioff-Merz05}.
More recently, Rahli~\etal formalised an executable version of
Multi-Paxos in EventML~\cite{Rahli-al:AVOCS15}, a dialect of NuPRL.  
Dragoi~\etal~\cite{Dragoi-al:POPL16} implemented and verified SD-Paxos in the
\textsc{PSync} framework, which implements a partially synchronised
model~\cite{CharronBost-Merz:IJSI09}, supporting automated proofs of system
invariants.
Padon~\etal have proved the system invariants and the consensus property of
both simple Paxos and \mpaxos using the verification tool
\textsc{Ivy}~\cite{Padon-al:PLDI16,Padon-al:OOPSLA17}.

Unlike all those verification efforts that consider
(Multi-/Disk/Fast/$\ldots$)Paxos as a \emph{single monolithic protocol}, our
approach provides the first \emph{modular} verification of single-decree Paxos
using Rely/Guarantee framework, as well as the first verification of \mpaxos
that directly reuses the proof of SD-Paxos.


\paragraph{Compositional Reasoning about Distributed Systems.}

Several recent works have partially addressed modular formal
verification of distributed systems.
The IronFleet framework by Hawblitzel~\etal has been used to verify
both safety and liveness of a real-world implementation of a
Paxos-based replicated state machine library and a lease-based shared
key-value store~\cite{Hawblitzel-al:SOSP15}. While the proof is
structured in a modular way by composing specifications in a way
similar to our decomposition in
Sections~\ref{sec:faithfull-deconstruction}--\ref{sec:nodet-specs},
that work does not address the linearisability and does not provide
composition of proofs about complex protocols (\eg, \mpaxos) from
proofs about its subparts

The Verdi framework for deductive verification of distributed
systems~\cite{Wilcox-al:PLDI15,Woos-al:CPP16} suggests the idea of
\emph{Verified System Transformers} (VSTs), as a way to provide
\emph{vertical composition} of distributed system implementation.
While Verdi's VSTs are similar in its purpose and idea to our
network transformations, they \emph{do not} exploit the properties of
the protocol, which was crucial for us to verify \mpaxos's
implementation.

The \textsc{Disel} framework~\cite{Sergey-al:POPL18,Wilcox-al:SNAPL17}
addresses the problem of \emph{horizontal composition} of distributed
protocols and their client applications. While we do not compose Paxos
with any clients in this work, we believe its register-based
specification could be directly employed for verifying applications
that use Paxos as its sub-component, which is what is demonstrated by
our prototype implementation.


\section{Conclusion and Future Work}
\label{sec:conclusion}

We have proposed and explored two complementary mechanisms for modular
verification of Paxos-family consensus protocols~\cite{Lamport:TOPLAS98}: (a)
non-deterministic register-based specifications in the style of
Boichat~\etal~\cite{Boichat-al:SN03}, which allow one to decompose the proof
of protocol's linearisability into separate independent ``layers'', and (b) a
family of protocol-aware transformations of network semantics, making it
possible to reuse the verification efforts.
We believe that the applicability of these mechanisms spreads beyond reasoning
about Paxos and its variants and that they can be used for verifying other
consensus protocols, such as Raft~\cite{Ongaro-Ousterhout:UATC14} and
PBFT~\cite{Castro-Liskov:OSDI99}. We are also going to employ network
transformations to verify implementations of Mencius~\cite{Mao-al:OSDI08}, and
accommodate more protocol-specific optimisations, such as implementation of
master leases and epoch numbering~\cite{Chandra-al:PODC07}.

\paragraph{Acknowledgements.}
We thank the ESOP~2018 reviewers for their feedback.
%
%
This work by was supported by ERC Starting Grant H2020-EU~714729 and EPSRC
First Grant EP/P009271/1.


\bibliographystyle{abbrv}
\bibliography{references,proceedings}

\appendix
\section{Proof Outline of Module \emph{Paxos}}
\label{sec:outline-paxos}

\begin{proof}[Theorem~\ref{th:paxos}]
  By the following proof of linearisation. The following predicates state the
  relation that connects the concrete with the abstract state and the
  invariant of Paxos:
\begin{displaymath}
  \begin{array}{rcl}
    \AbsP &\equiv& \absvP = \vRC \land (\absvP = \Undefined \lor \absvP \in \valsRC)\\
    \InvP &\equiv&
    \begin{array}[t]{l}
      \valsRC \subseteq \{v\mid \exists (i, b).\ \ptpb{i} =
      \texttt{($b$, $v$)}\}.
    \end{array}
  \end{array}
\end{displaymath}

We consider actions (\texttt{ProposeP1})
\begin{displaymath}
  \begin{array}{l}
    \absvP = \vP = \Undefined \land v \not= \Undefined \land
    v \in \valsRC = V \land{}\\
    I = \{i \mid \ptpb{i} = \texttt{(true, $v$)}\} \land
    (\bigwedge_{i\in I}\absresPb{i} = \Undefined)
    \\
    \leadsto\\
    \absvP = \vP = v \land \valsRC = V \land{}\\
    (\bigwedge_{i\in I}(\ptpb{i} = \texttt{(false, $v$)} \land \absresPb{i} = v)),
  \end{array}
\end{displaymath}
(\texttt{ProposeP2})
\begin{displaymath}
  \begin{array}{l}
    \absvP = \vP = v \land v \not=\Undefined \land
    I = \{i \mid \ptpb{i} = \texttt{(true, $v$)}\} \land{}\\
    (\bigwedge_{i\in I}\absresPb{i} = \Undefined)
    \\
    \leadsto\\
    \absvP = \vP = v \land
    (\bigwedge_{i\in I}(\ptpb{i} = \texttt{(false, $v$)} \land \absresPb{i} = v)),
  \end{array}
\end{displaymath}
(\texttt{ProposeP3})$_i$
\begin{displaymath}
  \begin{array}{l}
    \absvP = \vP = \Undefined \land v \not= \Undefined \land
    v = \ptpb{i} \land{}\\
    I = \{i \mid \ptpb{i} = \texttt{(true, $v$)}\} \land
    (\bigwedge_{i\in I}\absresPb{i} = \Undefined)
    \\
    \leadsto\\
    \absvP = \vP = v \land
    (\bigwedge_{i\in I}(\ptpb{i} = \texttt{(false, $v$)} \land \absresPb{i} = v)),
  \end{array}
\end{displaymath}
and (\texttt{ProposeP4})$_i$
\begin{displaymath}
  \begin{array}{l}
    \absvP = \vP = v \land v \not=\Undefined \land
    I = \{i \mid \ptpb{i} = \texttt{(true, $v$)}\} \land{}\\
    (\bigwedge_{i\in I}\absresPb{i} = \Undefined) \land
    v' \not= v \land v' \not= \Undefined \land{}\\
    \ptpb{i} = v' \land \absresPb{i} = \Undefined
    \\
    \leadsto\\
    \absvP = \vP = v \land
    (\bigwedge_{i\in I}(\ptpb{i} = \texttt{(false, $v$)} \land \absresPb{i} = v)) \land{}\\
    \ptpb{i} = \Undefined \land \absresPb{i} = v.
  \end{array}
\end{displaymath}

The guarantee relation for \texttt{proposeP(v0)} is
\begin{displaymath}
  \begin{array}{l}
    (\texttt{ProposeP1}) \cup (\texttt{ProposeP2}) \cup
    (\texttt{ProposeP3})_{\prid} \cup (\texttt{ProposeP4})_{\prid},
  \end{array}
\end{displaymath}
where $\prid$ is the process identifier of the proposer, and the rely relation
is
\begin{displaymath}
  \begin{array}{l}
    (\texttt{ProposeP1}) \cup (\texttt{ProposeP2}) \cup
    \bigcup_{i\not=\prid}((\texttt{ProposeP3})_{i} \cup (\texttt{ProposeP4})_{i}).
  \end{array}
\end{displaymath}

{\setlength\arraycolseptemp{0pt}
\begin{lstlisting}
(@\hl{val abs\verbund vP := undef;}@)
(@\hl{(bool $\times$ val) ptp[1..$n$] := undef;}@)
(@\hl{single bool abs\verbund resP[1..$n$] := undef;}@)
proposeP(val v0) {
  int k; bool res; val v;
  (@\hl{assume(!(v0 = undef));}@)
  (@\Asrtn{$\left\{
      \begin{array}{l}
        \ptpb{\prid} = \Undefined \land \AbsP \land \InvP
      \end{array}
    \right\}$}@)
  k := pid(); (@\hl{ptp[pid()] := (true, v0);}@)
  (@\Asrtn{$\left\{
      \begin{array}{l}
        \prid = k \land \AbsP \land \InvP
      \end{array}
    \right\}$}@)
  do {
    (@\Asrtn{$\left\{
        \begin{array}{l}
          \prid = ((\kay - 1) \mmod n) + 1 \land \AbsP \land \InvP
        \end{array}
      \right\}$}@)
    (@\hl{$\langle$}@) (res, v) := proposeRC(k, v0);
      (@\hl{if (res) \verblbrace}@)
        (@\hl{for (i := 1, i <= $n$, i++) \verblbrace}@)
          (@\hl{if (ptp[i] = (true, v)) \verblbrace~lin(i); ptp[i] := (false, v); \verbrbrace~\verbrbrace}@)
        (@\hl{if (!(v = v0)) \verblbrace~lin(pid()); ptp[pid()] := (false, v0); \verbrbrace~\verbrbrace}@)
      (@\Asrtn{$\left\{
          \begin{array}{l}
            (\begin{array}[t]{l}
               (\begin{array}[t]{l}
                 \res = \true \land \absvP = \absresPb{\prid} = \va)
               \end{array}
               {}\lor
               \begin{array}[t]{l}
                 \res = \false) \land{}
               \end{array}
             \end{array}\\
            \ptpb{\prid} = \texttt{(false, v)} \land \prid = k \mmod n \land \AbsP \land \InvP
          \end{array}
        \right\}$}@)
    (@\hl{$\rangle$}@)
    (@\Asrtn{$\left\{
        \begin{array}{l}
          (\begin{array}[t]{l}
             (\begin{array}[t]{l}
                \res = \true \land \absvP = \absresPb{\prid} = \va)
              \end{array}
             {}\lor
             \begin{array}[t]{l}
               \res = \false) \land{}
             \end{array}
           \end{array}\\
          \ptpb{\prid} = \texttt{(false, v)} \land \prid = k \mmod n \land \AbsP \land \InvP
        \end{array}
      \right\}$}@)
    k := k + (@$n$@);
    (@\Asrtn{$\left\{
        \begin{array}{l}
          (\begin{array}[t]{l}
             (\begin{array}[t]{l}
                \res = \true \land  \absvP = \absresPb{\prid} = \va)
              \end{array}
             {}\lor
             \begin{array}[t]{l}
               \res = \false) \land{}
             \end{array}
           \end{array}\\
          \ptpb{\prid} = \texttt{(false, v)} \land \prid = k \mmod n \land \AbsP \land \InvP
        \end{array}
      \right\}$}@)
  } while (!res);
  (@\Asrtn{$\left\{
      \begin{array}{l}
        \absvP = \absresPb{\prid} = \va \land{}\\
        \ptpb{\prid} = \texttt{(false, v)}
        \land \prid = k \mmod n \land \AbsP \land \InvP
      \end{array}
    \right\}$}@)
  return v; }
\end{lstlisting}}
\qed
\end{proof}

\section{Proof Outline of Module \emph{Round-Based Consensus}}
\label{sec:outline-round-based-consensus}

\begin{proof}[Theorem \ref{th:round-based-consensus}]
  By the following proof of linearisation. The following predicates state the
  abstract relation $\AbsRC$ between the concrete and the abstract state in
  the instrumented implementation of
  Figure~\ref{fig:spec-inst-imp-round-based-consensus}.
\begin{displaymath}
  \begin{array}{rcl}
    \AbsRC &\equiv&
    \begin{array}[t]{l}
    \absvRC = \vRR \land \absroundRC \leq \roundRR \land{}\\
    \absvalsRC = \valsRR \setminus \{\Undefined\}.
    \end{array}
  \end{array}
\end{displaymath}
Variables \vRR, \roundRR\ and \valsRR\ are respectively the decided value, the
round and the set of values from the module \emph{Round-Based
  Register}. Predicate $\AbsRC$ ensures that the abstract \absvRC\ and
concrete \vRR\ coincide, that the abstract round $\absroundRC$ is less or
equal than the concrete $\roundRR$, and that the abstract $\absvalsRC$
corresponds to the concrete $\valsRR$ minus $\Undefined$.


The following predicate states the invariant $\InvRC$ of Round-Based
Consensus.
\begin{displaymath}
  \begin{array}{rcl}
    \InvRC &\equiv& (\absvRC = \Undefined \lor
    (v \not= \Undefined \land \absvRC = v \land v \in \absvalsRC)).
  \end{array}
\end{displaymath}
The invariant ensures that either no value has been decided yet (\ie\
$\absvRC = \Undefined$), or otherwise a value $v\not=\Undefined$ has been
decided (\ie\ $\absvRC = v$) and the abstract \absvalsRC\ contains $v$.

Now we define the rely and guarantee relations. We consider the actions
(\texttt{ProposeRC1})$_{k}$
\begin{displaymath}
  \begin{array}{l}
    \absroundRC \leq k = \roundRR \land \absvRC = \Undefined
    \land \absvalsRC = V\\
    \leadsto\\
    \absroundRC = \roundRR = k \land \absvRC = v \land v \not= \Undefined
    \land{}\\
    \absvalsRC = V \cup \{v\} \land \absresRC = \texttt{(true, $v$)},
  \end{array}
\end{displaymath}
(\texttt{ProposeRC2})$_{k}$
\begin{displaymath}
  \begin{array}{l}
    \absroundRC \leq k = \roundRR \land \absvRC = v \land v \not= \Undefined
    \land{}\\ v \in \absvalsRC = V\\
    \leadsto\\
    \absroundRC = \roundRR = k \land \absvRC = v \land v \not= \Undefined
    \land \absvalsRC = V,
  \end{array}
\end{displaymath}
(\texttt{ProposeRC3})$_{k}$
\begin{displaymath}
  \begin{array}{l}
    \absroundRC = k \land \roundRR = k' \land k \leq k' \land
    \absvRC = \Undefined
    \land{}\\ \absvalsRC = V\\
    \leadsto\\
    \absroundRC = k \land \roundRR = k' \land \absvRC = \Undefined
    \land v \not= \Undefined \land{}\\
    \absvalsRC = V \cup \{v\},
  \end{array}
\end{displaymath}
and (\texttt{ReadRC})$_{k}$
\begin{displaymath}
  \roundRR \leq k \leadsto \roundRR = k.
\end{displaymath}

The guarantee relation for \texttt{proposeRC(k, v0)} is
\begin{displaymath}
  (\texttt{ProposeRC1})_{\kay} \cup (\texttt{ProposeRC2})_{\kay} \cup
  (\texttt{ProposeRC3})_{\kay} \cup (\texttt{ReadRC})_{\kay},
\end{displaymath}
and the rely relation is
\begin{displaymath}
  \bigcup_{(k\,\mmod\,n) \not= (\kay\,\mmod\,n)}
  ((\texttt{ProposeRC1})_{k} \cup (\texttt{ProposeRC2})_{k} \cup
  (\texttt{ProposeRC3})_{k} \cup (\texttt{Read})_k).
\end{displaymath}



The proof outline below helps to show that if $\AbsRel \land \Inv$ holds at
the beginning of the method invocation of the annotated program, then it also
holds at the end of the method invocation after the abstract operation has
been performed at the linearisation point, and that the abstract result
$\absresRC$ coincides with the result of the concrete method. It also helps to
show that the method ensures the guarantee relation, that is, the states
between each atomic operation are related by the guarantee condition.

{\setlength\arraycolsep{0pt}
\begin{lstlisting}
(@\hl{val abs\verbund vRC := undef;}@)
(@\hl{int abs\verbund roundRC := 0;}@)
(@\hl{set of val abs\verbund valsRC := \verblbrace\verbrbrace;}@)
proposeRC(int k, val v0) {
  (@\hl{single (bool $\times$ val) abs\verbund resRC := undef;}@)
  bool res; val v;
  (@\hl{assume(!(v0 = undef));}@)
  (@\hl{assume(pid() = ((k - 1) mod $n$) + 1;}@)
  (@\Asrtn{$\left\{
      \begin{array}{l}
        \prid = ((\kay - 1) \mmod n) + 1 \land \vzero \not= \Undefined \land
        \absresRC = \Undefined \land{}\\
        \AbsRC \land \InvRC
      \end{array}\right\}$}@)
  (@\hl{$\langle$}@) (res, v) := read(k);
    (@\hl{if (res = false) \verblbrace~linRC(undef, \verbund); \verbrbrace}@)
    (@\Asrtn{$\left\{
        \begin{array}{l}
          (\begin{array}[t]{l}
             (\begin{array}[t]{l}
                \res = \true \land \va = \vRR \land \vRR \not= \Undefined
                \land \kay = \roundRR)
              \end{array}\\
             {}\lor
             (\begin{array}[t]{l}
                \resR = \true \land \va \in \valsRR)
              \end{array}\\
             {}\lor
             (\begin{array}[t]{l}
                \res = \false \land
                \absresRC = \texttt{(false, \verbund)})) \land{}
              \end{array}
           \end{array}\\
          \prid = ((\kay - 1) \mmod n) + 1 \land \vzero \not= \Undefined \land
          \AbsRC \land \InvRC
        \end{array}\right\}$}@)
  (@\hl{$\rangle$}@)
  (@\Asrtn{$\left\{
      \begin{array}{l}
        (\begin{array}[t]{l}
           (\begin{array}[t]{l}
              \res = \true \land \va = \vRR \land \vRR \not= \Undefined
              \land \kay \leq \roundRR)
            \end{array}\\
           {}\lor
           (\begin{array}[t]{l}
              \res = \true \land \va \in \valsRR)
            \end{array}\\
           {}\lor
           (\begin{array}[t]{l}
              \res = \false \land
              \absresRC = \texttt{(false, \verbund)})) \land{}
            \end{array}
         \end{array}\\
        \prid = ((\kay - 1) \mmod n) + 1 \land \vzero \not= \Undefined \land
        \AbsRC \land \InvRC
      \end{array}\right\}$}@)
  if (res) {
    (@\Asrtn{$\left\{
        \begin{array}{l}
          (\begin{array}[t]{l}
             (\begin{array}[t]{l}
                \va = \vRR \land \vRR \not= \Undefined
                \land \kay \leq \roundRR)
              \end{array}
             {}\lor
             (\begin{array}[t]{l}
                \va \in \valsRR))\land{}
              \end{array}
           \end{array}\\
           \prid = ((\kay - 1) \mmod n) + 1 \land \vzero \not= \Undefined \land
          \AbsRC \land \InvRC
        \end{array}\right\}$}@)
    if (v = undef) {
      (@\Asrtn{$\left\{
          \begin{array}{l}
            \va = \Undefined \land
            \va \in \valsRR\land{}\\
            \prid = ((\kay - 1) \mmod n) + 1 \land \vzero \not= \Undefined \land
            \AbsRC \land \InvRC
          \end{array}\right\}$}@)
      v := v0;
      (@\Asrtn{$\left\{
          \begin{array}{l}
            \va = \vzero\land{}\\
            \prid = ((\kay - 1) \mmod n) + 1 \land \vzero \not= \Undefined \land
            \AbsRC \land \InvRC
          \end{array}\right\}$}@)
    }
    (@\Asrtn{$\left\{
        \begin{array}{l}
          (\begin{array}[t]{l}
             (\begin{array}[t]{l}
                \va = \vRR \land \kay \leq \roundRR)
              \end{array}
             {}\lor
             \begin{array}[t]{l}
               \va \in \valsRR
             \end{array}
             {}\lor
             \begin{array}[t]{l}
               \va = \vzero)\land{}
             \end{array}
           \end{array}\\
          \prid = ((\kay - 1) \mmod n) + 1 \land \va \not= \Undefined \land
          \AbsRC \land \InvRC
        \end{array}\right\}$}@)
    (@\hl{$\langle$}@) res := write(k, v);
      (@\hl{if (res) \verblbrace~linRC(v, true); \verbrbrace}@)
      (@\hl{else \verblbrace~linRC(v, false);~\verbrbrace}@)
      (@\Asrtn{$\left\{
          \begin{array}{l}
            (\begin{array}[t]{l}
               (\begin{array}[t]{l}
                  \res = \true \land \vRR = \va \land \kay = \absroundRC = \roundRR
                  \land{}\\
                  \absresRC = \texttt{(true, v)})
                \end{array}\\
               {}\lor
               (\begin{array}[t]{l}
                  \res = \false \land \absresRC = \false))\land{}
                \end{array}
             \end{array}\\
            \va \in \valsRC \land \prid = ((\kay - 1) \mmod n) + 1 \land
            \AbsRC \land \InvRC
          \end{array}\right\}$}@)
      (@\hl{$\rangle$}@)
    (@\Asrtn{$\left\{
        \begin{array}{l}
          (\begin{array}[t]{l}
             (\begin{array}[t]{l}
                \res = \true \land \vRR = \va \land \kay \leq \roundRR
                \land \absresRC = \texttt{(true, v)})
              \end{array}\\
             {}\lor
             (\begin{array}[t]{l}
                \res = \false \land \absresRC = \texttt{(false, \verbund)}))\land{}
              \end{array}
           \end{array}\\
          \va \in \valsRC \land \prid = ((\kay - 1) \mmod n) + 1 \land
          \AbsRC \land \InvRC
        \end{array}\right\}$}@)
    if (res) {
      (@\Asrtn{$\left\{
          \begin{array}{l}
            \vRR = \va \land \kay \leq \roundRR
            \land \absresRC = \texttt{(true, v)} \land{}\\
            \va \in \valsRC \land \prid = ((\kay - 1) \mmod n) + 1 \land
            \AbsRC \land \InvRC
          \end{array}\right\}$}@)
      return (true, v); } }
  (@\Asrtn{$\left\{
      \begin{array}{l}
        \absresRC = \texttt{(false, \verbund)} \land
        \prid = ((\kay - 1) \mmod n) + 1 \land \AbsRC \land \InvRC
      \end{array}\right\}$}@)
  return (false, _); }
\end{lstlisting}}
\qed
\end{proof}

\section{Proof Outline of Module \emph{Round-Based Register}}
\label{sec:outline-round-based-register}

\begin{proof}[Theorem~\ref{th:round-based-register}]
We use the predicates $\send$ and $\receive$ to represent the state of the
network. The predicate $\send(i, j, m)$ is true iff process $i$ sent message
$m$ to process $j$. The predicate $\receive(j, i, m)$ is true iff, in turn,
process $j$ received message $m$ from~$i$.



We introduce the following abbreviations for the state of the network:
{\setlength\arraycolseptemp{0pt}
\begin{displaymath}
  \begin{array}[t]{rcl}
    \send(i,j,\mathit{msg},m)&\equiv&
    \begin{array}[t]{l}
      m \in M \land m.\From=i \land m.\To = j \land{}\\
      m.\mcont = \mathit{msg}
    \end{array}\\
    \receive(j,i,\mathit{msg},m)&\equiv&
    \begin{array}[t]{l}
      m \in M \land m.\From=i \land m.\To = j \land {}\\
      m.\mcont = \mathit{msg} \land  m.\Active = \False
    \end{array}\\
    \reqRead(i,j,k,m)&{}\equiv{}&
    \begin{array}[t]{l}
      \send(i,j,\texttt{[RE, $k$]},m) \land{}\\
      \lnot(\exists (k',v).\ \send(j, i, \texttt{[ackRE, $k$, $v$, $k'$]},m))
    \end{array}\\
    \ackRead(j,i,k,v,k',m)&{}\equiv{}&
    \begin{array}[t]{l}
      \receive(j,i,\texttt{[RE, $k$]},m) \land{}\\
      \send(j,i,\texttt{[ackRE, $k$, $v$, $k'$]},m)
    \end{array}\\
    \reqWrite(i,j,k,v,m)&{}\equiv{}&
    \begin{array}[t]{l}
      \send(i,j,\texttt{[WR, $k$, $v$])},m) \land{}\\
      \lnot\send(j,i,\texttt{[ackWR, $k$]},m)
    \end{array}\\
    \ackWrite(j,i,k,v)&{}\equiv{}&
    \begin{array}[t]{l}
      \receive(j,i,\texttt{[WR, $k$, $v$]},m) \land{}\\
      \send(j,i,\texttt{[ackWR, $k$]},m)
    \end{array}\\
  \end{array}
\end{displaymath}}

In our proofs of linearisation we consider the following proof rules
\begin{mathpar}
  \inferrule*
  { }
  {\Asrtn{\{\lnot(\exists m'.\ m'=m \land \send(\prid,j,\mesg,m')) \land p\}}\\\\
    \texttt{send($j$, $\mesg$)}\\\\
    \Asrtn{\{\send(\prid,j,\mesg, m) \land p\}}}
  \\
  \inferrule*[right={\normalsize ,}]
  { }
  {\Asrtn{\{\lnot(\exists m'.\ m'=m \land
      \receive(\prid,i,\mesg,m')) \land p\}}\\
    \texttt{(i, msg) := receive()}\\
    \Asrtn{\{\send(i, \prid,\mesg,m) \land \receive(\prid,i,\mesg,m) \land \ii = i
    \land \texttt{msg} = \mesg \land p\}}}
\end{mathpar}
which are sound under the network semantics of
Section~\ref{sec:multi-paxos-via} and under the operational semantics of our
pseudo-code that we outline in Appendix~\ref{ap:encoding-SD-Paxos}.


The following abbreviations express the invariant that connects the auxiliary
variables \texttt{count\verbund r} and \texttt{count\verbund w} with the
cardinality of the corresponding quorums.

{\setlength\arraycolsep{0pt}
\begin{displaymath}
  \begin{array}[t]{rcl}
    \CountR(k) &{}\equiv{}&
    \begin{array}[t]{l}
    \countrb{k} ={}\\
      \qquad|\{j\mid \exists (k',v.m).\ \ackRead(j,((k-1)\mmod n)+1,k,v,k',m)\}|
    \end{array}\\
    \CountW(k) &{}\equiv{}&
    \begin{array}[t]{l}
      \countwb{k} ={}\\
      |\{j\mid \exists (v,m).\ \ackWrite(j,((k-1)\mmod n)+1,k,v,m)\}|
    \end{array}\\
    \Count(k) &{}\equiv{}& \CountR(k) \land \CountW(k)
  \end{array}
\end{displaymath}}

The following predicates state the abstract relation $\AbsRR$ between the
concrete and the abstract state in the instrumented implementation of
Figures~\ref{fig:instrumented-read-write} and \ref{fig:instrumented-acceptor}.

{\setlength\arraycolsep{0pt}
\begin{displaymath}
  \begin{array}[t]{rcl}
    \AbstractV(v,k) &{}\equiv{}&
    \begin{array}[t]{l}
      (\absvRR = v \not= \Undefined \land k = \absroundRR)\\
      \Longrightarrow
      \begin{array}[t]{l}
        \exists Q.\
        (\begin{array}[t]{l}
           |Q| \geq \lceil(n+1)/2\rceil \land{}\\
           \forall j\in Q.\ \exists m.\ \ackWrite(j,((k-1)\mmod n)+1,k,v,m))
         \end{array}
      \end{array}
    \end{array}\\
    \AbstractVals(v) &{}\equiv{}&
    \begin{array}[t]{l}
    v\in \absvalsRR\\
    \Longrightarrow
      (\begin{array}[t]{l}
         v = \Undefined\\
         {}\lor \exists(j,k,m).\ \reqWrite(((k-1) \mmod n)+1,j,k,v,m))
       \end{array}
    \end{array}\\
    \AbstractRound(k) &{}\equiv{}&
    \begin{array}[t]{l}
      \absroundRR = k = 0\\
      {}\lor
      (\begin{array}[t]{l}
      \absroundRR =
        \max\{k\mid
        \begin{array}[t]{l}
          \countwb{k} \geq \lceil(n+1)/2\rceil\\
          {}\lor \countrb{k} \geq \lceil(n+1)/2\rceil\})
        \end{array}
      \end{array}
    \end{array}\\
    \AbsRR &{}\equiv{}&
    \forall (v,k).\ (\AbstractV(v,k) \land \AbstractVals(v) \land \AbstractRound(k))
  \end{array}
\end{displaymath}}

The following predicates state the invariant $\InvRR$ of \emph{Round-Based
  Register}.
{\setlength\arraycolsep{0pt}
\begin{displaymath}
  \begin{array}{rcl}
    \Read(j,k) &{}\equiv{}&
    \begin{array}[t]{l}
      \texttt{$j$.r}=k> 0\\
      \Longrightarrow
      (\begin{array}[t]{l}
         \exists (v,m).\ \ackWrite(j,((k-1)\mmod n)+1,k,v,m)\\
         {}\lor \exists (k',m).\ \ackRead(j,((k-1)\mmod n)+1,k,v,k',m))
       \end{array}
    \end{array}\\
    \Val(j,v) &{}\equiv{}&
    \begin{array}[t]{l}
      \texttt{$j$.v}=v\not=\texttt{undef}\\
      \Longleftrightarrow \exists (k,m).\
                              \ackWrite(j,((k-1)\mmod n)+1,k,v,m)
    \end{array}\\
    \ProphROne(k) &{}\equiv{}&
    \begin{array}[t]{l}
      (\prophrb{k} = \texttt{($k'$, $v$)} \land
      \countrb{k} \geq \lceil(n+1)/2\rceil)\\
      \Longrightarrow \absresrb{k} = (\true,v)
    \end{array}\\
    \ProphRTwo(k) &{}\equiv{}&
    \begin{array}[t]{l}
      \prophrb{k} = \texttt{(false, \verbund)} \Longrightarrow \absresrb{k} = \Undefined
    \end{array}\\
    \ProphWOne(k) &{}\equiv{}&
    \begin{array}[t]{l}
      (\prophwb{k} = \true \land \countwb{k} \geq \lceil(n+1)/2\rceil)\\
      \Longrightarrow \absreswb{k} = \true
    \end{array}\\
    \ProphWTwo(k) &{}\equiv{}&
    \begin{array}[t]{l}
      (\prophwb{k} = \false \land \countwb{k} > 0)\\
      \Longrightarrow \absreswb{k} = \false
    \end{array}\\
    \ProphWThree(k) &{}\equiv{}&
    \begin{array}[t]{l}
      (\prophwb{k} = \false \land \countwb{k} = 0)\\
      \Longrightarrow \absreswb{k} = \Undefined
    \end{array}\\
    \Proph(k) &{}\equiv{}&
    \begin{array}[t]{l}
      \begin{array}[t]{ll}
        \ProphROne(k) \land \ProphRTwo(k) \land{}\\
        \ProphWOne(k) \land \ProphWTwo(k) \land \ProphWThree(k)
      \end{array}
    \end{array}\\
    \InvRR &{}\equiv{}&
    \forall (j,k,v).\
    (\begin{array}[t]{l}
       \texttt{$j$.r}\geq\texttt{$j$.w}\land
       \Read(j,k) \land{}\\
       \Val(j,v) \land \Count(k)\land \Proph(k)).
     \end{array}
  \end{array}
\end{displaymath}}

The proof of Theorem~\ref{th:round-based-register} involves two proofs of
linearisation for \texttt{read} and \texttt{write} respectively, and one proof
proving that the code of acceptors meets the invariant $\AbsRR\land\InvRR$.

Now we define the rely and guarantee relations. We consider three kinds of
actors in the system corresponding to each of the proofs: reader, writer, and
acceptor. We define first the guarantee relations for each of the actors, and
then we express the rely relation for each actor as a combination of the
guarantee relations of the other actors. Consider the actions
(\texttt{Send})$_{(i,j,\mesg)}$
\begin{displaymath}
  m.\From = i\land m.\To = j \land m.\mcont = \mesg \leadsto
  \send(i,j,\mesg,m),
\end{displaymath}
and (\texttt{Receive})$_{(j,i,\mesg)}$
\begin{displaymath}
  \begin{array}{l}
    m.\From = i\land m.\To = j \land m.\mcont = \mesg
    \land \lnot(\receive(j, i, \mesg,m))\\
    \leadsto \send(i,j,\mesg,m) \land \receive(j,i,\mesg,m),
  \end{array}
\end{displaymath}
which model sending and receiving a message.

A reader can also perform actions
(\texttt{ReadFails1})$_{k}$
\begin{displaymath}
  \begin{array}{l}
    \receive(i,j,\texttt{[nackRE', $k$])},m)
    \land i = ((k-1) \mmod n) + 1 \land{}\\
    \prophrb{k} = \texttt{(false, \verbund)} \land \absresrb{k} = \Undefined
    \land k \geq \absroundRR\\
    \leadsto\\
    \receive(i,j,\texttt{[nackRE', $k$]},m) \land
    \prophrb{k} = \Undefined \land{}\\
    \absresrb{k} = \texttt{(false, \verbund)} \land \absroundRR = k
  \end{array}
\end{displaymath}
and (\texttt{ReadFails2})$_{k}$
\begin{displaymath}
  \begin{array}{l}
    \receive(i,j,\texttt{[nackRE, $k$]},m)
    \land i = ((k-1) \mmod n) + 1 \land{}\\
    \prophrb{k} = \texttt{(false, \verbund)} \land \absresrb{k} = \Undefined
    \land k < \absroundRR = k'\\
    \leadsto\\
    \receive(i,j,\texttt{[nackRE, $k$]},m) \land
    \prophrb{k} = \Undefined \land{}\\
    \absresrb{k} = \texttt{(false, \verbund)} \land \absroundRR = k'.
  \end{array}
\end{displaymath}

The guarantee relation of \texttt{read($k$)} is the one induced by the union of these
actions as follows:
\begin{displaymath}
    (\Reader)_{k} \equiv
    \begin{array}[t]{l}
      \bigcup_{j}
      (\texttt{Send})_{((k\,\mmod\,n)+1,j,\texttt{[RE,$k$]})} \cup{}\\
      \bigcup_{j,v,k'}
      (\begin{array}[t]{l}
      (\texttt{Receive})_{((k\,\mmod\,n)+1,j,\texttt{[ackRE,$k$,$v$,$k'$]})} \cup{} \\
      (\texttt{Receive})_{((k\,\mmod\,n)+1,j,\texttt{[nackRE,$k$]})}) \cup{}
      \end{array}\\
      (\texttt{ReadFails1})_{k} \cup
      (\texttt{ReadFails2})_{k}.
    \end{array}
\end{displaymath}

Now we focus on a writer, which, additionally to sending reads and receiving
acknowledgements, can perform action (\texttt{WriteFails1})$_{(k,v)}$
\begin{displaymath}
  \begin{array}{l}
    \receive(i,j,\texttt{[nackWR, $k$]}, m)
    \land i = ((k-1) \mmod n)+1 \land v \not= \Undefined
    \land{}\\
    \prophwb{k} = \false \land \absvalsRR = V \land \countwb{k} = 0 \land{}\\
    \absreswb{k} = \Undefined\\
    \leadsto\\
    \receive(i,j,\texttt{[nackWR, $k$]}, m) \land i = (k \mmod n)+1
    \land{}\\
    \prophwb{k} = \Undefined \land \absvalsRR = V \cup \{v\} \land \countwb{k} = 0 \land{}\\
    \absreswb{k} = \false.
  \end{array}
\end{displaymath}

The guarantee relation of \texttt{write($k$, $v$)} is the one induced by the union
of these actions as follows:
\begin{displaymath}
    (\Writer)_{(k,v)} \equiv
    \begin{array}[t]{l}
      \bigcup_{j}
      (\texttt{Send})_{(((k-1)\,\mmod\,n)+1,j,\texttt{[WR,$k$,$v$]})} \cup{}\\
      \bigcup_{j}
      (\begin{array}[t]{l}
      (\texttt{Receive})_{(((k-1)\,\mmod\,n)+1,j,\texttt{[ackWR,$k$]})} \cup{} \\
      (\texttt{Receive})_{(((k-1)\,\mmod\,n)+1,j,\texttt{[nackWR,$k$]})}) \cup{}
      \end{array}\\
      (\texttt{WriteFails1})_{(k,v)}.
    \end{array}
\end{displaymath}

Now to the acceptors, which can receive read and write requests and send
non-acknowledgements to them. They can also perform actions
(\texttt{ReadSucceeds1})$_{j}$
\begin{displaymath}
  \begin{array}{l}
    \reqRead(i,j,k,m) \land i = ((k-1) \mmod n) + 1 \land k \geq \texttt{$j$.r} \land {}\\
    \absresrb{k} = \Undefined \land \prophrb{k} = \texttt{(true, $v$)}
    \land v \in \absvalsRR = V \land{}\\
    \countrb{k} = c = \lceil(n+1)/2\rceil - 1 \land  k \geq \absroundRR\\
    \leadsto\\
    \ackRead(j,i,k,\texttt{$j$.v},\texttt{$j$.w},m) \land \texttt{$j$.r} = k
    \land \countrb{k} = c + 1 \land{}\\
    \prophrb{k} = \texttt{(true, $v$)} \land \absroundRR = k
    \land \absvalsRR = V \land{}\\ \absresrb{k} = \texttt{(true, $v$)},
  \end{array}
\end{displaymath}
(\texttt{ReadSucceeds2})$_{j}$
\begin{displaymath}
  \begin{array}{l}
    \reqRead(i,j,k,m) \land i = ((k-1) \mmod n) + 1 \land k \geq \texttt{$j$.r} \land {}\\
    \absresrb{k} = \Undefined \land \prophrb{k} = \texttt{(true, $v$)}
    \land v \in \absvalsRR = V \land{}\\
    \countrb{k} = c = \lceil(n+1)/2\rceil - 1 \land  k < \absroundRR = r\\
    \leadsto\\
    \ackRead(j,i,k,\texttt{$j$.v},\texttt{$j$.w},m) \land \texttt{$j$.r} = k
    \land \countrb{k} = c + 1 \land{}\\
    \prophrb{k} = \texttt{(true, $v$)} \land \absroundRR = r
    \land \absvalsRR = V \land{}\\ \absresrb{k} = \texttt{(true, $v$)},
  \end{array}
\end{displaymath}
(\texttt{AckRead1})$_{j}$
\begin{displaymath}
  \begin{array}{l}
    \reqRead(i,j,k,m) \land i = ((k-1) \mmod n) + 1
    \land k \geq \texttt{$j$.r}\land {}\\
    \countrb{k} = c \not= \lceil(n+1)/2\rceil - 1\\
    \leadsto\\
    \ackRead(j,i,k,\texttt{$j$.v},\texttt{$j$.w},m) \land \texttt{$j$.r} = k \land
    \countrb{k} = c + 1,
  \end{array}
\end{displaymath}
(\texttt{AckRead2})$_{j}$
\begin{displaymath}
  \begin{array}{l}
    \reqRead(i,j,k,m) \land i = ((k-1) \mmod n) + 1
    \land k \geq \texttt{$j$.r}\land {}\\
    \countrb{k} = c \land \prophrb{k} = p \not= \texttt{(true, $v$)}\\
    \leadsto\\
    \ackRead(j,i,k,\texttt{$j$.v},\texttt{$j$.w},m) \land \texttt{$j$.r} = k \land
    \countrb{k} = c + 1 \land{}\\  \prophrb{k} = p,
  \end{array}
\end{displaymath}
(\texttt{AckRead3})$_{j}$
\begin{displaymath}
  \begin{array}{l}
    \reqRead(i,j,k,m) \land i = ((k-1) \mmod n) + 1
    \land k \geq \texttt{$j$.r} \land {}\\
    \countrb{k} = c \land \absresrb{k} = \mathit{ar} \not= \Undefined\\
    \leadsto\\
    \ackRead(j,i,k,\texttt{$j$.v},\texttt{$j$.w},m) \land \texttt{$j$.r} = k
    \land \countrb{k} = c + 1 \land{}\\
    \absresrb{k} = \mathit{ar},
  \end{array}
\end{displaymath}
(\texttt{WriteSucceeds})$_{j}$
\begin{displaymath}
  \begin{array}{l}
    \reqWrite(i,j,k,v,m) \land i = ((k-1) \mmod n) + 1 \land v \not= \Undefined
    \land k \geq \texttt{$j$.r} \land{}\\
    \absvalsRR = V \land
    \absreswb{k} = \Undefined \land \prophwb{k} = \true \land{}\\
    \countwb{k} = c = \lceil(n+1)/2\rceil - 1 \land  k \geq \absroundRR\\
    \leadsto\\
    \ackWrite(j,i,k,v,m) \land \texttt{$j$.r} = k \land \texttt{$j$.w} = k
    \land \texttt{$j$.v} = v \land{}\\
    \countwb{k} = c + 1 \land
    \prophwb{k} = \true \land \absvalsRR = V \cup \{v\} \land{}\\
    \absroundRR = k \land \absvRR = v
    \land \absreswb{k} = \true,
  \end{array}
\end{displaymath}
(\texttt{AckWrite1})$_{j}$
\begin{displaymath}
  \begin{array}{l}
    \reqWrite(i,j,k,v,m) \land i = ((k-1) \mmod n) + 1 \land v \not= \Undefined
    \land k \geq \texttt{$j$.r} = r \land{}\\
    \absreswb{k} = \Undefined \land \prophwb{k} = \true \land{}\\
    \countwb{k} = c \not= \lceil(n+1)/2\rceil - 1\\
    \leadsto\\
    \ackWrite(j,i,k,v,m) \land \texttt{$j$.r} = k \land \texttt{$j$.w} = k
    \land \texttt{$j$.v} = v \land{}\\
    \countwb{k} = c + 1 \land
    \prophwb{k} = \true \land
    \absreswb{k} = \Undefined,
  \end{array}
\end{displaymath}
(\texttt{WriteFails2})$_{j}$
\begin{displaymath}
  \begin{array}{l}
    \reqWrite(i,j,k,v,m) \land i = ((k-1) \mmod n) + 1 \land v \not= \Undefined
    \land k \geq
    \texttt{$j$.r} = r \land{}\\
    \absreswb{k} = \Undefined \land \prophwb{k} = \false
    \land v \in \absvalsRR = V \land{}\\
    \countwb{k} = c\\
    \leadsto\\
    \ackWrite(j,i,k,v,m) \land \texttt{$j$.r} = k \land \texttt{$j$.w} = k
    \land \texttt{$j$.v} = v \land\absvalsRR = V \cup \{v\} \land{}\\
    \countwb{k} = c + 1 \land
    \prophwb{k} = \false \land
    \absreswb{k} = \false,
  \end{array}
\end{displaymath}
(\texttt{AckWrite2})$_{j}$
\begin{displaymath}
  \begin{array}{l}
    \reqWrite(i,j,k,v,m) \land i = ((k-1) \mmod n) + 1 \land v \not= \Undefined
    \land k \geq
    \texttt{$j$.r} = r \land{}\\
    \absreswb{k} = \mathit{ar} \land \prophwb{k} = \Undefined \land \countwb{k} = c\\
    \leadsto\\
    \ackWrite(j,i,k,v,m) \land \texttt{$j$.r} = k \land \texttt{$j$.w} = k
    \land \texttt{$j$.v} = v \land{}\\
    \countwb{k} = c + 1 \land
    \prophwb{k} = \Undefined \land
    \absreswb{k} = \mathit{ar},
  \end{array}
\end{displaymath}
and (\texttt{AckWrite3})$_{j}$
\begin{displaymath}
  \begin{array}{l}
    \reqWrite(i,j,k,v,m) \land i = ((k-1) \mmod n) + 1 \land v \not= \Undefined
    \land k \geq
    \texttt{$j$.r} = r \land{}\\
    \absreswb{k} = \mathit{ar} \not= \Undefined \land \prophwb{k} = p
    \land \countwb{k} = c\\
    \leadsto\\
    \ackWrite(j,i,k,v,m) \land \texttt{$j$.r} = k \land \texttt{$j$.w} = k
    \land \texttt{$j$.v} = v \land{}\\
    \countwb{k} = c + 1 \land
    \prophwb{k} = p \land \absreswb{k} = \mathit{ar}.
  \end{array}
\end{displaymath}

The guarantee relation for the code of acceptor $j$ is the one induced by the
union of these actions as follows:
\begin{displaymath}
  \begin{array}{rcl}
    (\Acceptor)_j &\equiv&
    \begin{array}[t]{l}
      \bigcup_{k}
      (\begin{array}[t]{l}
         (\texttt{Receive})_{(j,((k-1)\,\mmod\,n)+1,\texttt{[RE,$k$]})} \cup{}\\
         (\texttt{Send})_{(((k-1)\,\mmod\,n)+1,j,\texttt{[nackRE,$k$]})}) \cup{}\\
         (\texttt{Send})_{(((k-1)\,\mmod\,n)+1,j,\texttt{[nackWR,$k$]})}) \cup{}
       \end{array}\\
      \bigcup_{k,v\not=\Undefined}
      ((\texttt{Receive})_{(j,((k-1)\,\mmod\,n)+1,\texttt{[WR,$k$,$v$]})}) \cup{}\\
      (\texttt{ReadSucceeds2})_j \cup
      (\texttt{AckRead1})_j \cup
      (\texttt{AckRead2})_j \cup{}\\
      (\texttt{AckRead3})_j \cup\\
      (\texttt{WriteSucceeds})_j \cup
      (\texttt{AckWrite1})_j \cup
      (\texttt{WriteFails2})_j \cup{}\\
      (\texttt{AckWrite2})_j \cup
      (\texttt{AckWrite3})_j.
    \end{array}
  \end{array}
\end{displaymath}

The rely relation for both \texttt{read(k)} and \texttt{write(k, vW)} is
\begin{displaymath}
  \begin{array}{l}
    \bigcup_j(\Acceptor)_j \cup
    \bigcup_{k\not=\kay}(\Reader)_{k} \cup
    \bigcup_{k\not=\kay,v\not=\Undefined}(\Writer)_{(k,v)},
  \end{array}
\end{displaymath}
and rely relation for the code of acceptor \texttt{j} is
\begin{displaymath}
  \begin{array}{l}
    \bigcup_{j\not=\jay}(\Acceptor)_j \cup
    \bigcup_{k}(\Reader)_{k} \cup
    \bigcup_{k,v\not=\Undefined}(\Writer)_{(k,v)}.
  \end{array}
\end{displaymath}

The proof outline below helps to show that if $\AbsRR \land \InvRR$ holds at
the beginning of a method invocation, for both \texttt{read(k)} and
\texttt{write(k,~vW)}, then it also holds and the end of the method invocation
after the abstract operation has been performed at the linearisation point,
and that the abstract result (\absresrb{k} and \absreswb{k} respectively)
coincide with the result of the concrete method. It also helps to show that
each method ensures the corresponding guarantee relation, this is, the states
between any atomic operation are in the guarantee relation.

{\setlength\arraycolsep{0pt}
\begin{lstlisting}
(@\hl{val abs\verbund vRR := undef;}@)
(@\hl{int abs\verbund round := 0;}@)
(@\hl{set of val abs\verbund valsRR := \verblbrace undef\verbrbrace;}@)
(@\hl{val abs\verbund res\verbund r[1..$\infty$] := undef;}@)
(@\hl{val abs\verbund res\verbund w[1..$\infty$] := undef;}@)
(@\hl{int count\verbund r[1..$\infty$] := 0;}@)
(@\hl{int count\verbund w[1..$\infty$] := 0;}@)
(@\hl{(bool $\times$ val) proph\verbund r[1..$\infty$] := undef;}@)
(@\hl{bool proph\verbund w[i..$\infty$] := undef;}@)

read(int k) {
  int j; val v; int kW; val maxV; int maxKW; set of int Q; msg m;
  (@\hl{assume(pid() = ((k - 1) mod $n$) + 1);}@)
  (@\Asrtn{$\left\{
      \prid = ((\kay - 1) \mmod n) + 1 \land \AbsRR \land \InvRR
    \right\}$}@)
  (@\hl{$\langle$ if (\textrm{operation reaches} PL: RE\verbund SUCC \textrm{and define} $v = \texttt{maxV}$ \textrm{at that time}) \verblbrace}@)
      (@\hl{proph\verbund r[k] := (true, $v$); \verbrbrace}@)
    (@\hl{else \verblbrace~if (\textrm{operation reaches} PL: RE\verbund FAIL) \verblbrace}@)
      (@\hl{proph\verbund r[k] := (false, \verbund); \verbrbrace~\verbrbrace~$\rangle$}@)
  (@\Asrtn{$\left\{
      \prid = ((\kay - 1) \mmod n) + 1 \land \AbsRR \land \InvRR
    \right\}$}@)
  for (j := 1, j <= (@$n$@), j++) { send(j, [RE, k]); }
  (@\Asrtn{$\left\{
      \prid = ((\kay - 1) \mmod n) + 1 \land \AbsRR \land \InvRR
    \right\}$}@)
  maxKW := 0; maxV := undef; Q := {};
  (@\Asrtn{$\left\{
      \begin{array}{l}
        \maxKW = 0 \land \maxV = \Undefined \land{}\\
        \countrb{\kay} \geq |\texttt{Q}| \land \prid = ((\kay - 1) \mmod n) + 1
        \land \AbsRR \land \InvRR
      \end{array}
    \right\}$}@)
  do {
    (@\Asrtn{$\left\{
        \begin{array}{l}
          \maxKW = \max(\{k'\mid \exists (j,v,m).\ j\in \texttt{Q} \land
          \ackRead(j,i,\kay,v,k',m)\}\cup \{0\}) \land{}\\
          (\maxKW = 0 \lor (\exists (j,m).\ j\in \texttt{Q} \land
          \ackRead(j,i,\kay,\maxV,\maxKW,m)) \land{}\\
          \countrb{\kay} \geq |\texttt{Q}| \land i = \prid = ((\kay - 1) \mmod n) + 1
          \land \AbsRR \land \InvRR
        \end{array}
      \right\}$}@)
    (j, m) := receive();
    (@\Asrtn{$\left\{
        \begin{array}{l}
          \send(\jay,i,\texttt{m},m) \land
          \receive(i,\jay,\texttt{m},m)
          \land{}\\
          \maxKW = \max(\{k'\mid \exists (j,v,m).\ j\in \texttt{Q} \land
          \ackRead(j,i,\kay,v,k',m)\}\cup \{0\}) \land{}\\
          (\maxKW = 0 \lor (\exists (j,m).\ j\in \texttt{Q} \land
          \ackRead(j,i,\kay,\maxV,\maxKW,m)) \land{}\\
          \countrb{\kay} > |\texttt{Q}| \land i = \prid = ((\kay - 1) \mmod n) + 1
          \land \AbsRR \land \InvRR
        \end{array}
      \right\}$}@)
    switch (m) {
      case [ackRE, @k, v, kW]:
        (@\Asrtn{$\left\{
            \begin{array}{l}
              \ackRead(\jay,i,\kay,\va,\kW,m) \land{}\\
              \maxKW = \max(\{k'\mid \exists (j,v,m).\ j\in \texttt{Q} \land
              \ackRead(j,i,\kay,v,k',m)\}\cup \{0\}) \land{}\\
              (\maxKW = 0 \lor (\exists (j,m).\ j\in \texttt{Q} \land
              \ackRead(j,i,\kay,\maxV,\maxKW,m)) \land{}\\
              \countrb{\kay} > |\texttt{Q}| \land i = \prid = ((\kay - 1) \mmod n) + 1
              \land \AbsRR \land \InvRR
            \end{array}
          \right\}$}@)
        Q := Q (@$\cup$@) {j};
        (@\Asrtn{$\left\{
            \begin{array}{l}
              \jay \in \texttt{Q} \land\ackRead(\jay,i,\kay,\va,\kW,m) \land{}\\
              \maxKW = \max(\{k'\mid \exists (j,v,m).\ j\in \texttt{Q} \land
              \ackRead(j,i,\kay,v,k',m)\}\cup \{0\}) \land{}\\
              (\maxKW = 0 \lor (\exists (j,m).\ j\in \texttt{Q} \land
              \ackRead(j,i,\kay,\maxV,\maxKW,m)) \land{}\\
              \countrb{\kay} \geq |\texttt{Q}| \land i = \prid = ((\kay - 1) \mmod n) + 1
              \land \AbsRR \land \InvRR
            \end{array}
          \right\}$}@)
        if (kW >= maxKW) {
          (@\Asrtn{$\left\{
              \begin{array}{l}
                \kW \geq \maxKW \land \jay \in \texttt{Q}
                \land\ackRead(\jay,i,\kay,\va,\kW,m) \land{}\\
                \maxKW = \max(\{k'\mid \exists (j,v,m).\ j\in \texttt{Q} \land
                \ackRead(j,i,\kay,v,k',m)\}\cup \{0\}) \land{}\\
                (\maxKW = 0 \lor (\exists (j,m).\ j\in \texttt{Q} \land
                \ackRead(j,i,\kay,\maxV,\maxKW,m)) \land{}\\
                \countrb{\kay} \geq |\texttt{Q}| \land i = \prid = ((\kay - 1) \mmod n) + 1
                \land \AbsRR \land \InvRR
              \end{array}
            \right\}$}@)
          maxKW := kW; maxV := v;
          (@\Asrtn{$\left\{
              \begin{array}{l}
                \maxKW = \kW \land \maxV = \va \land
                \jay \in \texttt{Q} \land\ackRead(\jay,i,\kay,\va,\kW,m) \land{}\\
                \maxKW = \max(\{k'\mid \exists (j,v,m).\ j\in \texttt{Q} \land
                \ackRead(j,i,\kay,v,k',m)\}\cup \{0\}) \land{}\\
                (\maxKW = 0 \lor (\exists (j,m).\ j\in \texttt{Q} \land
                \ackRead(j,i,\kay,\maxV,\maxKW,m)) \land{}\\
                \countrb{\kay} \geq |\texttt{Q}| \land i = \prid = ((\kay - 1) \mmod n) + 1
                \land \AbsRR \land \InvRR
              \end{array}
            \right\}$}@)
        }
        (@\Asrtn{$\left\{
            \begin{array}{l}
              \jay \in \texttt{Q} \land\ackRead(\jay,i,\kay,\va,\kW,m) \land{}\\
              \maxKW = \max(\{k'\mid \exists (j,v,m).\ j\in \texttt{Q} \land
              \ackRead(j,i,\kay,v,k',m)\}\cup \{0\}) \land{}\\
              (\maxKW = 0 \lor (\exists (j,m).\ j\in \texttt{Q} \land
              \ackRead(j,i,\kay,\maxV,\maxKW,m)) \land{}\\
              \countrb{\kay} \geq |\texttt{Q}| \land i = \prid = ((\kay - 1) \mmod n) + 1
              \land \AbsRR \land \InvRR
            \end{array}
          \right\}$}@)
      case [nackRE, @k]:
        (@\Asrtn{$\left\{
            \begin{array}{l}
              \prophrb{\kay} = \texttt{(false, \verbund)}
              \land \absresrb{\kay} = \Undefined \land{}\\
              \maxKW = \max(\{k'\mid \exists (j,v,m).\ j\in \texttt{Q} \land
              \ackRead(j,i,\kay,v,k',m)\}\cup \{0\}) \land{}\\
              (\maxKW = 0 \lor (\exists (j,m).\ j\in \texttt{Q} \land
              \ackRead(j,i,\kay,\maxV,\maxKW,m)) \land{}\\
              \countrb{\kay} \geq |\texttt{Q}| \land i = \prid = ((\kay - 1) \mmod n) + 1
              \land \AbsRR \land \InvRR
            \end{array}
          \right\}$}@)
        (@\hl{$\langle$ linRE(k, undef, false); proph\verbund r[k] := undef; }@)
          return (false, _); (@\hl{$\rangle$ // PL: RE\verbund FAIL}@)
    }
    (@\Asrtn{$\left\{
        \begin{array}{l}
          \jay \in \texttt{Q} \land\ackRead(\jay,i,\kay,\va,\kW,m) \land{}\\
          \maxKW = \max(\{k'\mid \exists (j,v,m).\ j\in \texttt{Q} \land
          \ackRead(j,i,\kay,v,k',m)\}\cup \{0\}) \land{}\\
          (\maxKW = 0 \lor (\exists (j,m).\ j\in \texttt{Q} \land
          \ackRead(j,i,\kay,\maxV,\maxKW,m)) \land{}\\
          \countrb{\kay} \geq |\texttt{Q}| \land i = \prid = ((\kay - 1) \mmod n) + 1
          \land \AbsRR \land \InvRR
        \end{array}
      \right\}$}@)
    if ((@$|\texttt{Q}|$@) = (@$\lceil$@)(n+1)/2(@$\rceil$@)) {
      (@\Asrtn{$\left\{
          \begin{array}{l}
            \prophrb{\kay} = \texttt{(true, maxV)} \land
            \jay \in \texttt{Q} \land\ackRead(\jay,i,\kay,\va,\kW,m) \land{}\\
            \maxKW = \max(\{k'\mid \exists (j,v,m).\ j\in \texttt{Q} \land
            \ackRead(j,i,\kay,v,k',m)\}\cup \{0\}) \land{}\\
            (\maxKW = 0 \lor (\exists (j,m).\ j\in \texttt{Q} \land
            \ackRead(j,i,\kay,\maxV,\maxKW,m)) \land{}\\
            \countrb{\kay} \geq |\texttt{Q}| \land i = \prid = ((\kay - 1) \mmod n) + 1
            \land \AbsRR \land \InvRR
          \end{array}
        \right\}$}@)
      return (true, maxV); (@\hl{// PL: RE\verbund SUCC}@)
    }
    (@\Asrtn{$\left\{
        \begin{array}{l}
          \maxKW = \max(\{k'\mid \exists (j,v,m).\ j\in \texttt{Q} \land
          \ackRead(j,i,\kay,v,k',m)\}\cup \{0\}) \land{}\\
          (\maxKW = 0 \lor (\exists (j,m).\ j\in \texttt{Q} \land
          \ackRead(j,i,\kay,\maxV,\maxKW,m)) \land{}\\
          \countrb{\kay} \geq |\texttt{Q}| \land i = \prid = ((\kay - 1) \mmod n) + 1
          \land \AbsRR \land \InvRR
        \end{array}
      \right\}$}@)
  } while (true); }

write(int k, val vW) {
  int j; set of int Q; msg m;
  (@\hl{assume(!(vW = undef));}@)
  (@\hl{assume(pid() = ((k - 1) mod $n$) + 1);}@)
  (@\Asrtn{$\left\{
      \prid = ((\kay - 1) \mmod n) + 1 \land \vW \not= \Undefined \land \AbsRR \land \InvRR
    \right\}$}@)
  (@\hl{$\langle$ if (\textrm{operation reaches} PL: WR\verbund SUCC) \verblbrace~proph\verbund w[k] := true; \verbrbrace}@)
    (@\hl{else \verblbrace~if (\textrm{operation reaches} PL: WR\verbund FAIL) \verblbrace}@)
              (@\hl{proph\verbund w[k] := false; \verbrbrace~\verbrbrace~$\rangle$}@)
  (@\Asrtn{$\left\{
      \prid = ((\kay - 1) \mmod n) + 1 \land \vW \not= \Undefined \land \AbsRR \land \InvRR
    \right\}$}@)
  for (j := 1, j <= (@$n$@), j++) { send(j, [WR, k, vW]); }
  (@\Asrtn{$\left\{
      \prid = ((\kay - 1) \mmod n) + 1 \land \vW \not= \Undefined \land \AbsRR \land \InvRR
    \right\}$}@)
  Q := {};
  (@\Asrtn{$\left\{
      \countwb{\kay} \geq |\texttt{Q}| \land \prid = ((\kay - 1) \mmod n) + 1
      \land \vW \not= \Undefined \land \AbsRR \land \InvRR
    \right\}$}@)
  do {
    (@\Asrtn{$\left\{
        \countwb{\kay} \geq |\texttt{Q}| \land \prid = ((\kay - 1) \mmod n) + 1
        \land \vW \not= \Undefined \land
        \AbsRR \land \InvRR
      \right\}$}@)
    (j, m) := receive();
    (@\Asrtn{$\left\{
        \begin{array}{l}
          \send(\jay,i,\texttt{m}) \land \receive(i,\jay,\texttt{m}) \land
          \countwb{\kay} \geq |\texttt{Q}| \land{}\\ i = \prid = ((\kay - 1) \mmod n) + 1
          \land \vW \not= \Undefined \land
          \AbsRR \land \InvRR
        \end{array}
      \right\}$}@)
    switch (m) {
      case [ackWR, @k]:
        (@\Asrtn{$\left\{
            \begin{array}{l}
              \ackWrite(\jay,i,\kay,\vW) \land
              \countwb{\kay} > |\texttt{Q}| \land{}\\ i = \prid = ((\kay - 1) \mmod n) + 1
              \land \vW \not= \Undefined \land \AbsRR \land \InvRR
            \end{array}\right\}$}@)
        Q := Q (@$\cup$@) {j};
        (@\Asrtn{$\left\{
            \begin{array}{l}
              \ackWrite(\jay,i,\kay,\vW) \land
              \countwb{\kay} \geq |\texttt{Q}| \land{}\\ i = \prid = ((\kay - 1) \mmod n) + 1
              \land \vW \not= \Undefined \land \AbsRR \land \InvRR
            \end{array}\right\}$}@)
      case [nackWR, @k]:
        (@\Asrtn{$\left\{
            \begin{array}{l}
              \prophrb{\kay} = \texttt{fail}\land
              \countwb{\kay} \geq |\texttt{Q}| \land{}\\ \prid = ((\kay - 1) \mmod n) + 1
              \land \vW \not= \Undefined \land \AbsRR \land \InvRR
            \end{array}\right\}$}@)
        (@\hl{$\langle$ if (count\verbund w[k] = 0) \verblbrace}@)
            (@\hl{linWR(k, vW, false); proph\verbund w[k] := undef; \verbrbrace}@)
          return false; (@\hl{$\rangle$ // PL: WR\verbund FAIL}@)
    }
    (@\Asrtn{$\left\{
        \countwb{\kay} \geq |\texttt{Q}| \land \prid = ((\kay - 1) \mmod n) + 1
        \land \vW \not= \Undefined \land \AbsRR \land \InvRR
      \right\}$}@)
    if ((@$|\texttt{Q}|$@) = (@$\lceil$@)(n+1)/2(@$\rceil$@)) {
      (@\Asrtn{$\left\{
          \begin{array}{l}
            \prophrb{\kay} = \texttt{true} \land
            \countwb{\kay} \geq \lceil (n+1)/2\rceil \land{}\\
            \prid = ((\kay - 1) \mmod n) + 1 \land \vW \not= \Undefined \land \AbsRR \land \InvRR
          \end{array}
        \right\}$}@)
      return true; (@\hl{// PL: WR\verbund SUCC}@)
    }
    (@\Asrtn{$\left\{
        \countwb{\kay} \geq |\texttt{Q}| \land \prid = ((\kay - 1) \mmod n) + 1
        \land \vW \not= \Undefined \land \AbsRR \land \InvRR
      \right\}$}@)
  } while (true); }
\end{lstlisting}}

The proof outline below helps to show that the code of each process acceptor
meets the invariant $\AbsRR \land \InvRR$ and also ensures the guarantee
relation, this is, the states between any atomic operation are in the
corresponding guarantee relation.

{\setlength\arraycolsep{0pt}
\begin{lstlisting}
process Acceptor(int j) {
  val v := undef; int r := 0; int w := 0;
  start() {
    int i; msg m; int k;
    do {
      (@\Asrtn{$\left\{
          \prid = \jay \land \AbsRR \land \InvRR
        \right\}$}@)
      (i, m) := receive();
      (@\Asrtn{$\left\{
          \begin{array}{l}
            \send(\ii,\jay,\texttt{m}, m) \land
            \receive(\jay,\ii,\texttt{m}, m) \land{}\\
            \ii = ((\kay - 1) \mmod n) + 1 \land \prid = \jay \land \AbsRR \land \InvRR
          \end{array}
        \right\}$}@)
      switch (m) {
        case [RE, k]:
          (@\Asrtn{$\left\{
              \reqRead(\ii,\jay,\kay, m) \land \ii = ((\kay - 1) \mmod n) + 1
              \land \prid = \jay \land \AbsRR \land \InvRR
            \right\}$}@)
          if (k < r) {
            (@\Asrtn{$\left\{
                \begin{array}{l}
                  \kay < \ar \land \reqRead(\ii,\jay,\kay, m) \land{}\\
                  \ii = ((\kay - 1) \mmod n) + 1 \land
                  \prid = \jay \land \AbsRR \land \InvRR
                \end{array}
              \right\}$}@)
            send(i, [nackRE, k]);
            (@\Asrtn{$\left\{
                \begin{array}{l}
                  \send(\jay, \ii, \texttt{(`nackRE', k, \verbund, \verbund)}, m)
                  \land \kay < \ar \land \reqRead(\ii,\jay,\kay, m) \land{}\\
                  \ii = ((\kay - 1) \mmod n) + 1 \land
                  \prid = \jay \land \AbsRR \land \InvRR
                \end{array}
              \right\}$}@)
          }
          else {
            (@\Asrtn{$\left\{
                \begin{array}{l}
                  \kay \geq \ar \land \reqRead(\ii,\jay,\kay, m) \land{}\\
                  \ii = ((\kay - 1) \mmod n) + 1 \land
                  \prid = \jay \land \AbsRR \land \InvRR
                \end{array}
              \right\}$}@)
            (@$\langle$@) r := k;
              (@\hl{if (abs\verbund res\verbund r[k] = undef) \verblbrace}@)
                (@\hl{if (proph\verbund r[k] = (true, $v$)) \verblbrace}@)
                  (@\hl{if (count\verbund r[k] = $\lceil$(n+1)/2$\rceil$ - 1) \verblbrace}@)
                    (@\hl{linRE(k, $v$, true); \verbrbrace~\verbrbrace~\verbrbrace}@)
              (@\hl{count\verbund r[k]++;}@) send(i, [ackRE, k, v, w]);
              (@\Asrtn{$\left\{
                  \begin{array}{l}
                    (\begin{array}[t]{l}
                       (\begin{array}[t]{l}
                          \countrb{\kay} = \lceil(n+1)/2\rceil \land
                          \prophrb{\kay} = \texttt{(true, $v$)} \land{}\\
                          \absroundRR \leq \kay \land \absvRR = v \land
                          \absresrb{\kay} = \texttt{(true, $v$)})
                       \end{array}\\
                       {}\lor
                       (\begin{array}[t]{l}
                          (\countrb{\kay} \not= \lceil(n+1)/2\rceil \lor
                          \prophrb{\kay} \not = \texttt{(true, \verbund)}) \land{}\\
                          \absresrb{\kay} = \Undefined))
                        \end{array}\\
                       {}\lor \absresrb{\kay} \not= \Undefined)\land{}\\
                       \countrb{\kay} > 0 \land \ar = \kay \land
                       \ackRead(\jay,\ii,\kay,\va,\dblu,m) \land{}\\
                       \ii = ((\kay - 1) \mmod n) + 1 \land
                       \prid = \jay \land \AbsRR \land \InvRR
                     \end{array}
                  \end{array}
                \right\}$}@)
            (@$\rangle$@)
          }
          (@\Asrtn{$\left\{
              \prid = \jay \land \AbsRR \land \InvRR
            \right\}$}@)
        case [WR, k, vW]:
          (@\Asrtn{$\left\{
              \begin{array}{l}
                \reqWrite(\ii,\jay,\kay,\vW, m) \land \vW \not= \Undefined
                \land{}\\
                \ii = ((\kay - 1) \mmod n) + 1 \land
                \prid = \jay \land\AbsRR \land \InvRR
              \end{array}
            \right\}$}@)
          if (k < r) {
            (@\Asrtn{$\left\{
                \begin{array}{l}
                  \kay < \ar \land \reqWrite(\ii,\jay,\kay,\vW, m)
                  \land \vW \not= \Undefined \land{}\\
                  \ii = ((\kay - 1) \mmod n) + 1 \land
                  \prid = \jay \land\AbsRR \land \InvRR
                \end{array}
              \right\}$}@)
            send(j, i, [nackWR, k]);
            (@\Asrtn{$\left\{
                \begin{array}{l}
                  \send(\jay, \ii, \texttt{[nackWR, k]},m)
                  \land \kay < \ar \land \reqWrite(\ii,\jay,\kay,\vW,m) \land{}\\
                  \vW \not= \Undefined \land \ii = ((\kay - 1) \mmod n) + 1 \land
                  \prid = \jay \land\AbsRR \land \InvRR
                \end{array}
              \right\}$}@)
          }
          else {
            (@\Asrtn{$\left\{
                \begin{array}{l}
                  \kay \geq \ar \land \reqWrite(\ii,\jay,\kay,\vW, m)
                  \land \vW \not= \Undefined \land{}\\
                  \ii = ((\kay - 1) \mmod n) + 1 \land
                  \prid = \jay \land\AbsRR \land \InvRR
                \end{array}
              \right\}$}@)
            (@$\langle$@) r := k; w := k; v := vW;
              (@\hl{if (abs\verbund res\verbund w[k] = undef) \verblbrace}@)
                (@\hl{if (!(proph\verbund w[k] = undef)) \verblbrace}@)
                  (@\hl{if (proph\verbund w[k]) \verblbrace}@)
                    (@\hl{if (count\verbund w[k] = $\lceil$(n+1)/2$\rceil$ - 1) \verblbrace}@)
                      (@\hl{linWR(k, vW, true); \verbrbrace~\verbrbrace}@)
                  (@\hl{else \verblbrace~linWR(k, vW, false); \verbrbrace~\verbrbrace~\verbrbrace}@)
              (@\hl{count\verbund w[k]++;}@) send(j, i, [ackWR, k]);
              (@\Asrtn{$\left\{
                  \begin{array}{l}
                    (\begin{array}[t]{l}
                       (\begin{array}[t]{l}
                          \countwb{\kay} = \lceil(n+1)/2\rceil \land
                          \prophwb{\kay} = \true \land{}\\
                          \absroundRR \geq \kay \land \absvRR = \vW \land
                          \absreswb{\kay} = \true)
                        \end{array}\\
                       {}\lor
                       (\begin{array}[t]{l}
                          \countwb{\kay} \not= \lceil(n+1)/2\rceil \land
                          \prophwb{\kay} = \true \land{}\\
                          \absreswb{\kay} = \Undefined)
                        \end{array}\\
                       {}\lor
                       (\begin{array}[t]{l}
                          \prophwb{\kay} = \false \land
                          \absreswb{\kay} = \false)
                        \end{array}\\
                       {}\lor \prophwb{\kay} = \Undefined
                       \lor \absreswb{\kay} \not= \Undefined) \land{}
                     \end{array}\\
                    \countwb{\kay} > 0 \land
                    \ar = \dblu = \kay \land \va = \vW \land
                    \ackWrite(\jay,\ii,\kay,\vW, m) \land{}\\
                    \vW \not= \Undefined \land \ii = ((\kay - 1) \mmod n) + 1 \land
                    \prid = \jay \land \AbsRR \land \InvRR
                  \end{array}
                \right\}$}@)
            (@$\rangle$@)
          }
          (@\Asrtn{$\left\{
              \prid = \jay \land \AbsRR \land \InvRR
            \right\}$}@)
      }
      (@\Asrtn{$\left\{
          \prid = \jay \land \AbsRR \land \InvRR
        \right\}$}@)
    } while (true); }
}
\end{lstlisting}}
\qed
\end{proof}

\section{Encoding SD-Paxos as an Abstract Protocol}
\label{ap:encoding-SD-Paxos}

Let $\Prog$ be the set of programs of a language that subsumes the imperative
while language for our pseudo-code in
Sections~\ref{sec:faithfull-deconstruction} and \ref{sec:nodet-specs}, and
which adds a parallel composition operator $\parallel$, which is commutative
and associative, and a null process $0$, which is the neutral element of
$\parallel$. The semantics of the parallel composition operator that we need
here is very simplistic and it does not pose any issue regarding the
interaction of the components within the parallel composition. (The
interaction will be implemented on top of this operational semantics by the
network semantics of the abstract distributed protocols introduced in
Section~\ref{sec:multi-paxos-via}.) The only purpose of $\parallel$ here is to
have processes that adopt both the roles of acceptor and proposer, and to
allow any of these two roles to make a move. The language $\Prog$ is morally a
sequential programming language.

We let $\Nodes = \mathbb{N}$ be the set of natural numbers, and $\mvoc.\mcont$
be the set of contents of the messages in the message vocabulary $\mvoc$,
which contains requests for read and write, and their corresponding
acknowledgements and non-acknowledgements as described in
Section~\ref{sec:faithfull-deconstruction}.

Now we assume a small-step operational semantics à la Winskel \cite{Win93} for
the programs in $\Prog$. We let the relation
$\opstepi{i} : (\Prog \times \Delta_\nod) \times (\Prog \times \Delta_\nod)$
be given by the operational semantics of a program run by node $i$ whose
current program line is not any of the network operations \texttt{send} or
\texttt{receive} and where $\Delta_\nod$ is the set of local states. We fix
relations
$\opsteps{i} : (\Prog \times \Delta_\nod) \times (\Prog \times \Delta_\nod
\times \mvoc)$
and
$\opstepr{i} : (\Prog \times \Delta_\nod \times \mvoc) \times (\Prog \times
\Delta_\nod)$ to be the ones induced by the rules:
\begin{mathpar}
  \inferrule*[lab=Send,width=5cm]
  {
    P = (\texttt{send($J$,\,$C$);\,$P_1$})\parallel P_2 \\
    P' = P_1 \parallel P_2 \\   m.\To = J \\
    m.\mcont=C\\ m.\From = i \\
  }
  {
    \langle P, \delta\rangle \opsteps{i}
    \langle P', \delta, m\rangle
  }
  \and
  \inferrule*[lab=Receive,width=7cm]
  {
    P = (\texttt{($J$,\,$C$) := receive();\,$P_1$})\parallel P_2 \\
    P' = P_1 \parallel P_2 \\
    m.\mcont=c  \\ m.\From = j \\ m.\To = i\\
    \delta' = \delta[J\mapsto j,C\mapsto c]\\
  }
  {
    \langle P, \delta, m\rangle \opstepr{i}
    \langle P', \delta'\rangle
  }
\end{mathpar}
The $J$ and the $C$ in rule \textsc{Send} above are meta-variables for some
expressions of $\Prog$ with types $\Nodes$ and $\mvoc.\mcont$ respectively. In
rule \textsc{Receive}, there is an assignment and the $J$ and the $C$ this
time are meta-variables for names of some fields in the local state
$\delta\in\Delta_\nod$, which we assume have types $\Nodes$ and $\mvoc.\mcont$
respectively (\eg, in our implementation of SD-Paxos in
Figure~\ref{fig:implementation-rb-register-read-write}, $J$ and $C$ are
respectively substituted by \texttt{j} and \texttt{[RE,\,k]} in line~5, and by
\texttt{j} and \texttt{m} in line~8).

Next, we will fix a particular set of local states $\Delta_\nod$ that matches
with our implementation of SD-Paxos in
Section~\ref{sec:faithfull-deconstruction}, and we will derive an abstract
distributed protocol for SD-Paxos from the relations defined in the previous
paragraph. We write $\mathbb{B} = \{\True, \False\}$ for the set of Booleans
and $\mathbb{V}$ for the set of values that are decided by Paxos. We
distinguish two sets of local state, $\Delta_\acc$ and $\Delta_\pro$, for
acceptors and proposers respectively:
\begin{displaymath}
  \begin{array}{rcl}
    \Delta_\acc&=& \{\texttt{j}:\Nodes;\, \texttt{v}:\mathbb{V};\,
                   \texttt{r}:\mathbb{N};\, \texttt{w}:\mathbb{N};\,
                   \texttt{i}:\Nodes;\,\texttt{m}:\mvoc.\mcont;\,
                   \texttt{k}:\mathbb{N}\} \\[4pt]
    \Delta_\pro&=&
    {\setlength\arraycolsep{0pt}
      \{\begin{array}[t]{l}
        \texttt{v0}:\mathbb{V};\, \texttt{kP}:\mathbb{N};\,
        \texttt{resP}:\mathbb{B};\, \texttt{vP}:\mathbb{V};\,
        \texttt{resRC}:\mathbb{B};\, \texttt{vRC}:\mathbb{V};\,\\
        \texttt{jRR}:\Nodes;\, \texttt{QRR}:\{\Nodes\};\,
        \texttt{mRR}:\mvoc.\mcont;\,
        \texttt{vRE}:\mathbb{V};\, \texttt{kW}:\mathbb{N};\,\\
        \texttt{maxV}:\mathbb{V};\, \texttt{maxKW}:\mathbb{N}\}
      \end{array}}
  \end{array}
\end{displaymath}

A local state for an acceptor $\delta\in\Delta_\acc$ contains a copy of the
parameters and fields of process \texttt{Acceptor} and the local variables of
its task \texttt{start()} in
Figure~\ref{fig:implementation-rb-register-acceptor} of
Section~\ref{sec:faithfull-deconstruction}. A local state for a proposer
$\delta\in\Delta_\pro$ contains a copy of the parameters and the local
variables of the client code in
Figures~\ref{fig:implementation-rb-register-read-write} and
\ref{fig:implementation-paxos-rb-consensus} of
Section~\ref{sec:faithfull-deconstruction}. For simplicity, we flatten the
code of \texttt{proposeP} on the right of
Figure~\ref{fig:implementation-paxos-rb-consensus} by inlining the codes of
\texttt{proposeRC}, \texttt{read} and \texttt{write}. To avoid clashing names,
we have appended one of the suffixes \texttt{P}, \texttt{RC}, \texttt{RR} or
\texttt{RE} to the names of some of the variables, and we have used top-most
variables instead of parameters of inlined methods when possible. For
instance, fields \texttt{vP}, \texttt{vRC} and \texttt{vRE} correspond
respectively to the variable \texttt{v} in each of the methods
\texttt{proposeP}, \texttt{proposeRC} and \texttt{read}, field \texttt{v0} is
used in place of the variable with the same name in both \texttt{proposeP} and
\texttt{proposeRC}, field \texttt{kP} is used in place of the variable
\texttt{k} in every method, and field \texttt{vRC} is used in place of the
variable \texttt{vW} in \texttt{write}. We have also reused fields
\texttt{jRR}, \texttt{QRR} and \texttt{mRR} in place of variables \texttt{j},
\texttt{Q} and \texttt{m} in both \texttt{read} and \texttt{write}, since
these two methods are invoked sequentially and do not interfere with each
other.

For each acceptor $i$ with local state $\delta\in \Delta_\acc$, we tag the
node by letting $\delta.\role=\mathit{Acceptor}$ and we let the implicit filed
$\delta.\pid$ coincide with $i$ and with the field
$\delta.\texttt{j}$. Customarily, all the nodes are acceptors. If, besides,
$i$ is a proposer that proposes value $v$, then its local state $\delta$ is in
$\Delta_\acc\times\Delta_\pro$ and we tag the node
$\delta.\role=\mathit{Proposer}$ and additionally we let
$\delta.\texttt{v0} = v$.  The set of local states for the operational
semantics of our implementation of SD-Paxos is
\begin{displaymath}
  \Delta_\nod = (\Delta_\acc + (\Delta_\acc \times \Delta_\pro))
\end{displaymath}

Now we can fix the start configurations for the operational semantics. For
each acceptor $i$ that is not a proposer, the initial local state is
\begin{displaymath}
  \delta_\acc^i[\texttt{j}\mapsto i, \texttt{v}\mapsto \bot,\texttt{r}\mapsto 0,
  \texttt{w}\mapsto 0]\in\Delta_\acc
\end{displaymath}
and the relation $\opstepi{i}$ starts at the configuration
$\langle\texttt{start()}\parallel 0,\delta_\acc^i\rangle$, where
\texttt{start()} is the code of an acceptor in
Figure~\ref{fig:implementation-rb-register-acceptor} and $0$ is the null
process.

For each proposer $i$ that proposes value $v$, the initial local state is
\begin{displaymath}
  \delta_\pro^i[\texttt{j}\mapsto i, \texttt{v}\mapsto \bot,\texttt{r}\mapsto 0,
  \texttt{w}\mapsto 0, \texttt{v0}\mapsto v]\in\Delta_\acc\times\Delta_\pro
\end{displaymath}
and the relation $\opstepi{i}$ starts at the configuration
$\langle\texttt{start()}\parallel\texttt{proposeP},\delta_\pro^i\rangle$,
where \texttt{proposeP} is the client code obtained by flattening the codes in
Figures~\ref{fig:implementation-rb-register-read-write} and
\ref{fig:implementation-paxos-rb-consensus} by inlining \texttt{proposeRC},
\texttt{read} and \texttt{write}, as explained in the previous paragraphs.

Now we can define an \emph{SD-Paxos sequence of a node $i$} as a sequence of
configurations $\langle P_0,\delta_0\rangle$, $\langle P_1,\delta_1\rangle$,
\ldots\ in $\Prog\times\Delta_\nod$ such that
\begin{itemize}
\item {\rm
    $\langle P_0,\delta_0\rangle = \langle\texttt{start()}\parallel
    0,\delta_\acc^i\rangle$}
  if the node $i$ is an acceptor that is not a proposer, or otherwise {\rm
    $\langle P_0,\delta_0\rangle = \langle\texttt{start()}\parallel
    \texttt{proposeP},\delta_\pro^i\rangle$} if node $i$ is a proposer, and
\item for every $n$ then either {\rm
    $\langle P_n,\delta_n\rangle \opstepi{i} \langle
    P_{n+1},\delta_{n+1}\rangle$},
  or there exists a message $m$ such that {\rm
    $\langle P_n,\delta_n,m\rangle \opsteps{i} \langle
    P_{n+1},\delta_{n+1}\rangle$}
  or {\rm
    $\langle P_n,\delta_n\rangle \opstepr{i} \langle
    P_{n+1},\delta_{n+1},m\rangle$}.
\end{itemize}

The encoding of SD-Paxos as an abstract distributed protocol is code-aware,
this is, the text of the program that a particular process is running is
included in the local state. The text of the program provides information
about the program flow, and saves us from using auxiliary machinery to
represent control-related aspects. The local state for the abstract
distributed protocol is
\begin{displaymath}
  \Delta = \Prog \times \Delta_\nod
\end{displaymath}
We let $\sigma : \Nodes \rightharpoonup \Delta$ be such that for each process
identifier $i$ in the range of $1$ to $n$, the local state of node $i$ is
$\delta = \sigma(i)$. Notice that the implicit field $\delta.\pid$ coincides
with the origin $i$ of the mapping $[i \mapsto \delta] \in \sigma$.

Now we can define the relations $\stepi$, $\steps$ and $\stepr$ in terms of
an SD-Paxos sequence of a node.

A pair $\langle\delta,\delta'\rangle$ is in $\stepi$ iff either
$\delta=\delta'$, or otherwise there exist $i$ such that $\delta$ and
$\delta'$ are consecutive steps in some SD-Paxos sequence of node $i$ and
$\delta \opstepi{i} \delta'$.

A triple $\langle\delta,\delta',\{m\}\rangle$ is in $\steps$ iff there exist
$i$ such that $\delta$ and $\delta'$ are consecutive steps in some SD-Paxos
sequence of node $i$ and
$\langle \delta,m\rangle \opsteps{i}\delta'$.

A triple $\langle\delta,m,\delta'\rangle$ is in $\stepr$ iff there exist $i$
such that $\delta$ and $\delta'$ are consecutive steps in some SD-Paxos
sequence of node $i$ and
$\delta \opstepr{i}\langle\delta',m\rangle$.


\section{Proofs of Lemmas and Theorems in Section~\ref{sec:multi-paxos-via}}

\noindent
\textbf{Lemma~\ref{lm:otaref} (OTA refinement).}
\emph{  $\beh{\opstep{p} \cup \opstepota{p,\predota}} \subseteq \beh{{p}}$, where
  $p$ is an instance of the module Paxos, as defined in
  Section~\ref{sec:faithfull-deconstruction} and in
  Example~\ref{ex:simple-paxos}}.
  \begin{proof}
  By definition of the protocol in Example~\ref{ex:simple-paxos} a read request does not
  change a proposer's local state and does not have a non-trivial
  precondition.\qed
\end{proof}

\noindent
\textbf{Lemma~\ref{lm:sem-sim1} (Slot-replicating simulation).}  \emph{For all
  $I, i \in I$, $\beh{{\cartp}}|_{i} = \beh{{p}}$.}
\begin{proof}
  Both ways: by induction on the length of the target behaviour, taking into
  the account that we can always add ``stuttering'' states.\qed
\end{proof}

\noindent
\textbf{Lemma~\ref{lm:wsem}.}
\emph{If $T$ from \rulename{WStepReceiveT} is OTA-compliant with predicate
$\predota$, such that
$\beh{\opstep{p} \cup \opstepota{p,\predota}} \subseteq
\beh{\opstep{p}}$ and $p$ is $\predota$-monotone,
then $\beh{\opstep{\cartw}} \subseteq \beh{\opstep{\cartp}}$.
}%
\begin{proof}
  The proof is by induction on the length of the trace. The
  rule~\rulename{WStepReceiveT} can replicate a message $m \in T$ for multiple
  slots $j \in I$. To match adding a replica of $m$ for each $j$ with the
  corresponding per-slot executions of the simple semantics, we use
  Definitions~\ref{def:otacomp} and~\ref{def:pota-mono} to show that for any
  $j \in I$ we can relate its replica of $m$ to the current local state (due
  to $p$'s $\predota$-monotonicity). This allows us to emulate this step by a
  per-slot execution of the $\rulename{OTASend}$. The result then follows from the
  refinement assumption.\qed
\end{proof}

\noindent
\textbf{Theorem~\ref{thm:grand-theorem}.} \emph{The implementation of \mpaxos
  that uses a register provider and bunching network semantics refines the
  specification in Figure~\ref{fig:spec-multi-paxos}}.
  \begin{proof}
    The theorem is a straightforward consequence of the linearisation result
    stated by Theorem~\ref{th:paxos} and of the the results about the
    transformations of the network semantics stated by
    Lemmas~\ref{lm:otaref} to \ref{lm:bunch}.\qed
\end{proof}


\end{document}